\documentclass[preprint]{aastex}

\defcitealias{Jorgensen07}{J07} 	  	 
\defcitealias{Jorgensen08}{J08} 	  	 
\defcitealias{Jorgensen06}{J06}
\defcitealias{Enoch08young}{E09}
\defcitealias{Enoch08lifetimes}{E08}

\usepackage{graphicx,rotating,multirow,lscape,longtable,gensymb,arydshln}
\usepackage{natbib}
\newcommand{\Msun}{\mbox{M$_{\odot}$}}
\newcommand{\negExp}{\mbox{$^{-1}$}}

 \title{The Mass Distributions of Starless and Protostellar Cores in Gould Belt Clouds}
\author{Sarah I. Sadavoy\altaffilmark{1,2}, James Di Francesco\altaffilmark{2,1}, Sylvain Bontemps\altaffilmark{3}, S. Thomas Megeath\altaffilmark{4}, Luisa M. Rebull\altaffilmark{5}, Erin Allgaier\altaffilmark{4}, Sean Carey\altaffilmark{5}, Robert Gutermuth\altaffilmark{6}, Joe Hora\altaffilmark{7}, Tracy Huard\altaffilmark{8}, Caer-Eve McCabe\altaffilmark{5}, James Muzerolle\altaffilmark{9}, Alberto Noriega-Crespo\altaffilmark{5}, Deborah Padgett\altaffilmark{5}, Susan Terebey\altaffilmark{10}}
\altaffiltext{1}{Department of Physics and Astronomy, University of Victoria, PO Box 355, STN CSC, Victoria BC CANADA, V8W 3P6}
\altaffiltext{2}{National Research Council Canada, Herzberg Institute of Astrophysics, 5071 West Saanich Road, Victoria BC CANADA, V9E 2E7}
\altaffiltext{3}{Observatoire de Bordeaux, Site de Floirac, 2 rue de l`Observatoire, 33270 Floirac, FRANCE}
\altaffiltext{4}{Ritter Observatory, MS-113, University of Toledo, 2801 W. Bancroft St., Toledo, OH 43606, USA}
\altaffiltext{5}{Spitzer Science Center, Mail Stop 220-6, California Institute of Technology, 1200 East California Boulevard, Pasadena, CA 91125}
\altaffiltext{6}{University of Massachusetts, Smith College, Northampton, MA 01063}
\altaffiltext{7}{Harvard-Smithsonian Center for Astrophysics, 60 Garden Street, MS-42, Cambridge, MA 02138, USA}
\altaffiltext{8}{Department of Astronomy, University of Maryland, College Park, MD 20742, USA}
\altaffiltext{9}{Space Telescope Science Institute, 3700 San Martin Dr., Baltimore, MD 21218, USA}
\altaffiltext{10}{Department of Physics and Astronomy PS315, 5151 State University Drive, California State University at Los Angeles, Los Angeles, CA 90032, USA}
\email{ssadavoy@uvic.ca}

\begin{document}

\begin{abstract}

Using data from the SCUBA Legacy Catalogue (850 $\mu$m) and \emph{Spitzer Space Telescope} (3.6 - 70 $\mu$m), we explore dense cores in the Ophiuchus, Taurus, Perseus, Serpens, and Orion molecular clouds.  We develop a new method to discriminate submillimeter cores found by SCUBA as starless or protostellar, using point source photometry from Spitzer wide field surveys. First, we identify infrared sources with red colors associated with embedded young stellar objects (YSOs). Second, we compare the positions of these YSO-candidates to our submillimeter cores. With these identifications, we construct new, self-consistent starless and protostellar core mass functions (CMFs) for the five clouds. We find best fit slopes to the high-mass end of the CMFs of $-1.26 \pm 0.20$, $-1.22 \pm 0.06$, $-0.95 \pm 0.20$, and $-1.67 \pm 0.72$ for Ophiuchus, Taurus, Perseus, and Orion, respectively. Broadly, these slopes are each consistent with the $-1.35$ power-law slope of the Salpeter initial mass function (IMF) at higher masses, but suggest some differences. We examine a variety of trends between these CMF shapes and their parent cloud properties, potentially finding a correlation between the high-mass slope and core temperature. We also find a trend between core mass and effective size, but we are very limited by sensitivity. We make similar comparisons between core mass and size with visual extinction (for $A_V \ge 3$) and find no obvious trends. We also predict the numbers and mass distributions of cores that future surveys with SCUBA-2 may detect in each of these clouds.
\end{abstract}


\section{Introduction}\label{Intro}

The origin of stellar mass is not well understood. We know stars form in very cold and dense regions of gas and dust, deeply embedded within molecular clouds, but the dominant processes that guide the formation of small scale structure and the collapse of those structures into stars is unclear. We define zones of relatively high density on the small scales in molecular clouds as ``cores,'' which have masses $\sim$ \Msun\ within a radius of $\sim 0.07$ pc and would form a single star or a stellar system with a few stars (\citealt{difran07}). Thus, molecular cloud cores represent the cumulative result of all physical processes that organize cloud mass into relatively high densities on small scales.

The masses of molecular cloud cores are important probes of the initial conditions of star formation, and the relationships between these cores and any stellar products may be key to understanding the origin of stellar mass. Previous studies of these populations (for examples, see \citealt{Motte98}; \citealt{Nutter07}; \citealt{Ward-T07}; \citealt{Enoch08lifetimes}) have revealed that core mass functions (CMFs) roughly follow lognormal shapes that closely resemble the observed stellar initial mass function (IMF). In particular, these studies have shown great similarity between the high-mass slopes of CMFs and the high mass slope of the IMF. The CMF, however, may still depend on physical conditions within the parent cloud. Indeed, the relationship between the CMF and IMF is likely complex and a variety of factors, such as fragmentation (e.g., \citealt{Dobbs05}), competitive accretion (e.g., \citealt{Bonnell04}), turbulence (e.g., \citealt{Elmegreen02}), magnetic fields (e.g., \citealt{Shu04}), and radiative feedback (e.g., \citealt{Offner09}) may be relevant. 

The continuity between CMFs and the stellar IMF suggests that stellar mass is related to how material in molecular clouds is first collected into stellar precursors. Unfortunately, previous studies generally examined small regions within larger clouds (e.g., the L1688 region of Ophiuchus in \citealt{Motte98}) and over different wavelengths (e.g., 850 $\mu$m, 1.1 mm). As well, each group had developed a different set of conditions to identify small scale structure in molecular clouds, thereby these studies follow different numbers of cores, even for the same region (e.g., for Perseus, \citealt{Hatchell05}; \citealt{Kirk06}). In spite of the inconsistent analyses, these previous studies seem to produce CMFs that resemble the IMF shape. Nevertheless, it is difficult to compare CMFs from different studies and test for differences between clouds. It would be advantageous to look at core populations in various clouds with different physical conditions using a self-consistent analysis.

In addition, observations of dense cores over the last decade have revealed populations of cores with and without embedded young stars (\citealt{difran07}). Cores that contain a young stellar object (YSO) will have lost some of the surrounding material to outflows or to accretion onto the central body (\citealt{Myers08}). Also, their intrinsic temperatures may differ, biasing estimates of their masses. Thus, to obtain an accurate CMF, starless cores must be differentiated from protostellar cores, ie., those containing YSOs. 

Distinguishing between starless and protostellar cores depends on detecting an infrared source embedded in the dense material. Deeply embedded sources can be very faint, however, making such detections difficult. As well, many observed infrared sources are background galaxies or bright giant stars in our Galaxy, which must be identified as external to the cloud. Therefore, multi-band mid- to far-infrared photometry has emerged as an effective means for identifying and classifying embedded YSOs from external sources of infrared emission (e.g., \citealt{Allen04}; \citealt{Megeath04}; \citealt{Harvey06}; \citealt{Gutermuth08};  \citealt{Megeath09}; \citealt{Evans09}; \citealt{Rebull09}). 

Already, numerous studies have developed techniques to identify protostellar core populations from starless core populations in molecular clouds (e.g., \citealt{Jorgensen07}; \citealt{Enoch08young}). These techniques, however, employ very different constraints in infrared colors, magnitudes, or locations of YSO candidates with respect to the cores. As such, the CMFs produced by each technique will contain different biases, making comparisons between these studies unclear. Ideally, we would want to use the same constraints to compare CMFs over a large sample of molecular clouds to explore differences and similarities in core populations with environment.

Here, we use 850 $\mu$m data from the Submillimeter Common-User Bolometer Array (SCUBA) and just-released Spitzer data with a new self-consistent method of classifying cores to identify starless and protostellar cores in five clouds (Ophiuchus, Taurus, Perseus, Serpens, and Orion). With this discrimination, we produce consistent CMFs across the five clouds and thus, we can directly compare these distributions. In \S \ref{CloudsSection}, we discuss our sample choices, including the target clouds, the infrared and submillimeter data used in this study, and the individual core populations and selection criteria. In \S \ref{ResultsSection}, we discuss our new core classification technique and compare our results with two other previously developed methods. In \S \ref{DiscussionSection}, we examine the CMFs produced from our own classification method, and we compare these to standard formulations of the IMF. Also, we examine trends in the CMFs between the clouds, compare core properties with their surrounding environments, and make predictions as to what forthcoming instruments will detect. In \S \ref{conc}, we summarize our results.

\section{Clouds} \label{CloudsSection}
\subsection{Cloud Properties}

Our analysis focused on the Ophiuchus, Taurus, Perseus, Serpens, and Orion molecular clouds. These clouds were chosen due to a wealth of available data (e.g., millimeter, infrared) and their close proximity (all $< 500$ pc), resulting in maps with good linear resolution. Such small scale observations are necessary to resolve cores as well as to probe the physical properties and structure inside cores (\citealt{Ward-T07}). The five clouds studied here represent a variety of physical environments. For example, the Taurus cloud is undergoing only low-mass star formation (\citealt{Hartmann00}) whereas the Orion region contains active regions of massive star formation and is found within a large OB association (\citealt{PetersonMegeath08}). We found total cloud masses using our visual extinction maps (see \S \ref{2massDataSection} and Figure \ref{complete}). Assuming that 1 magnitude of extinction corresponds to $10^{21}$ molecules of hydrogen gas (1 $A_V = 10^{21}\ H_2$ cm$^{-2}$) and a mean molecular weight of 2.33, we measured the mass of the cloud via,
\begin{equation}
M (M_{\odot}) = 56.4\left(\frac{\theta}{\mbox{deg}}\right)^2\left(\frac{d}{\mbox{100 pc}}\right)^2\sum{A_{V,i}}
\end{equation}
where $\theta$ is the angular size of the extinction map pixels, $d$ is the cloud distance (listed in Table \ref{cloudMass}), and $A_{V,i}$ is the extinction in a given pixel element. Table \ref{cloudMass} gives the mass measurements we obtained for each cloud and the area over which we measured the mass. We constrained our cloud mass and size measurements to the regions with $A_V > 1$. For Serpens, the cloud is located close to the Galactic Plane, making its boundary less certain. Thus, we are likely including extinction from the nearby Auriga cloud and over-estimating its mass.

\subsection{Core Properties}\label{coreProp}

For each cloud, we assumed the core dust temperatures, $T_d$, and the 850 $\mu$m opacities, $\kappa_{850}$, were constant. Temperatures and opacities of core dust probably deviate within a given region due to different circumstances, such as proximity to an embedded cluster. Table \ref{prop} lists the assumed values for $T_d$ and Table \ref{cloudMass} gives the assumed values of distance for each cloud. 
 
We note that the temperatures listed in Table \ref{prop} were obtained via different techniques. \citet{Friesen09} and \citet{Rosolowsky08} derived kinetic temperatures, $T_K$, of dense gas in Ophiuchus and Perseus, respectively, using ammonia hyperfine structure lines, and we assume these temperatures are similar to $T_d$ since the densities in the cores are expected to be high (\citealt{Goldsmith01}). 
For the L1544 region in Taurus, \citet{Andre00} used spectral energy distribution (SED) fitting from ISO, SCUBA, and IRAM observations to obtain $T_d = 13$ K.
As part of the COMPLETE survey, \citet{Schnee05} used the $60\mu$m$/100 \mu$m flux density ratio for Serpens (see their Figure 5) to estimate its dust temperature.
For Orion, \citet{Wilson99} observed ammonia and derived rotational temperatures of $\sim$ 20 K. As with Ophiuchus and Perseus, we assume the rotational temperature is similar to the dust temperature since the gas traced by ammonia is dense ($\sim 10^4$ cm$^{-3}$). 

Clearly a common origin of $T_d$ would be preferable, but note that the masses of all cores in the CMFs will scale with $T_d$ and to first order the CMF shape will not depend on $T_d$ (see \S \ref{DiscussionSection} for further discussion; we assume a 30\% error in $T_d$ to derive uncertainties in our masses and CMFs). Self-consistent determinations of $T_d$ for cores in these clouds will soon be possible through SED fitting of $75-500\ \mu$m data from the Herschel Gould Belt Survey (\citealt{AndreSaraceno05}; \citealt{Ward-T07GBS}).

\subsection{Data}\label{DataSection}

We obtained our data from large-scale surveys and included wavelengths from the submillimeter (850 $\mu$m) to the infrared ($\sim 2\ \mu$m to $70\ \mu$m). We used the submillimeter data to probe the densest regions of each cloud and the infrared data to identify embedded protostars through emission and characterize the extended cloud structure through extinction. We discuss each of these data sets in turn below.

\subsubsection{SCUBA Maps}\label{ScubaDataSection}

We obtained our submillimeter data from the SCUBA Legacy Catalogue (SLC; see \citealt{difran08} for more details), which is an archive of all 850 $\mu$m and 450 $\mu$m mapping observations from SCUBA on the James Clerk Maxwell Telescope (JCMT). These observations generally focused on the high extinction regions within each cloud, thus the SLC data are patchy.
The SLC includes two sub-catalogues: the Fundamental Catalogue, which contains only objects identified from data with high quality atmospheric corrections (consisting of $\sim$ 78 \% of map data with an areal coverage of $\sim$ 19.6 deg$^2$), and the Extended Catalogue, which includes all the data regardless of quality (areal coverage of $\sim$ 29.3 deg$^2$). For consistency, the SLC used the algorithms developed by \citet{Johnstone00matrix} to similarly reduce these data. As well, the SLC used the 2D Clumpfind algorithm (\citealt{Clumpfind}) to identify structures in the continuum emission. Clumpfind identifies flux peaks over a certain noise level and then uses flux contours (down to a minimum threshold level) to assign boundaries to the cores and measure total fluxes (see \citealt{difran08}). For a discussion on the biases associated with different clump-finding algorithms, see \citet{CurtisRicher09}. 

To identify our submillimeter cores, we only used the 850 $\mu$m data, since the 450 $\mu$m observations have a greater absolute flux uncertainty than the 850 $\mu$m data by over a factor of two (\citealt{difran08}). Since we are more interested in accurate core fluxes than wide areal coverage of the clouds, we drew our samples from the Fundamental Catalogue. Table \ref{pixels} lists the cloud areal coverage observed by each of the surveys, where the second column indicates the area mapped in the Fundamental Catalogue.

Initially, the SLC set a Clumpfind threshold level equal to a factor 3 above the local noise level in a given map. For each of these objects, a second set of flux and size measurements, known as the ``alternative'' flux and radius, were calculated assuming a common threshold of 90 mJy beam\negExp, which is a factor 3 larger than the typical 850 $\mu$m noise level of all SLC maps, 30 mJy beam\negExp\ (\citealt{difran08}). For our core properties, we considered the ``alternative'' flux and radius measurements given in the SLC to provide a consistent mass sensitivity for all the cores in a given cloud. Generally, the two flux and size measurements were quite similar.

\subsubsection{Spitzer Space Telescope Maps}\label{SpitzerDataSection}

The \emph{Spitzer} Space Telescope (\citealt{Spitzer}) has two instruments that observe mid- and far-infrared wavelengths: IRAC (Infrared Array Camera; \citealt{IRAC}) at 3.6, 4.5, 5.8, and 8.0 $\mu$m and MIPS (Multiband Imaging Photometer for Spitzer; \citealt{MIPS}) at 24, 70, and 160 $\mu$m. With the high infrared sensitivity provided by Spitzer, these cameras provided excellent data for determining the presence of a protostar within highly extincted regions like cores. We used ``Molecular Cores to Planet Forming Disks'' (c2d; \citealt{Evans03,Evans09}) Legacy Project data for the Ophiuchus, Perseus, and Serpens molecular clouds (see \citealt{Jorgensen06}; \citealt{Harvey06}; \citealt{Rebull07}; \citealt{Harvey07long,Harvey07short}; \citealt{Padgett08}). The Taurus data came from Guest Observer (GO) observations for the Taurus Legacy Project (Padgett et al. 2010 in preparation), and the Orion data are from Guaranteed Time Observations (GTO) for the Spitzer Orion Survey  (Megeath et al. 2010 in preparation).

For the c2d clouds, Taurus, and most Orion observations, IRAC and MIPS observations were conducted over two epochs to identify and remove asteroids. Because the observations were conducted over two epochs at least 6 hours apart, even distant asteroids can be removed from the images and subsequent catalogues. Asteroids are prominent at 24 $\mu$m and 8 $\mu$m and recognizable in the other bands. Taurus, in particular, is located towards the ecliptic plane, and about half of the point sources detect at 24 $\mu$m in Taurus are asteroids (see Padgett et al. 2010 in preparation). 

For the c2d catalogue, the MIPS integration times were 3 seconds per sky pointing, with a given position observed 5 times for a total of 15 seconds. While the 24 $\mu$m and 70 $\mu$m bands covered roughly the same area of the sky, the 24 $\mu$m scans had longer total integration times (30 seconds) than the 70 $\mu$m scans (15 seconds; \citealt{Young05}). Observations at 160 $\mu$m were not included in the final c2d catalogues, since these data were affected by saturation and a large beam size (\citealt{Evans09}). Similar to MIPS, IRAC observed each cloud twice, with the two epochs separated by about 6.5 hours. The integration time per pointing with IRAC was 12 seconds (\citealt{Porras07}).

The Taurus observations were conducted and described by Padgett et al. (2008\footnote{\footnotesize Delivery Document}, 2010 in preparation), and consisted of IRAC and MIPS observations over a wide area (see Table \ref{pixels}). While the MIPS observations were obtained in the same fashion as the c2d maps, the IRAC maps sacrificed redundancy for area and consisted of only two (12-second) frames per position. At the distance of Taurus (140 pc), however, these shallow IRAC maps should still detect legitimate YSOs down to very low masses, unless these sources are exceptionally embedded.

For Orion, the MIPS observations are a combination of several different surveys with different observing modes. The MIPS observations consisted of both fast (17$\arcsec$ s\negExp) and medium (6.5$\arcsec$ s\negExp) scanning modes with total integration times ranging from 30 - 40 seconds. Similar to the c2d IRAC observations, several regions in Orion were observed with IRAC in high dynamic range mode, though some were observed only once in a faster frame time. Only the IRAC observations for Orion were made over two epochs to minimize artifacts and maximize four band coverage. For more information, see Megeath et al. (2010, in preparation).

The areal coverage of our five clouds by Spitzer are listed in column 3 of Table \ref{pixels}. Generally, MIPS and IRAC scans did not cover identical regions, though there was significant  overlap. Due to these different areal coverages and different effective sensitivities among the instruments, there are a number of sources that were detected at only a few wavelengths. For objects that were identified in the IRAC or MIPS maps but without valid photometry at a particular band, the c2d team ``band-filled'' the missing band by using a wavelength-appropriate point spread function (PSF) located at the expected position of the object. In cases where the pixels are compromised (e.g., from saturation or cosmic ray contamination) or the object is undetected, however, this process can result in negative band-filled fluxes. For more information, see the c2d Delivery Document, Evans et al. (2007). The Taurus and Orion data are not band-filled.

\subsubsection{2MASS Extinction Maps}\label{2massDataSection}

Infrared observations can probe the total line-of-sight column density of dust by tracing the color of stars located beyond a molecular cloud that are reddened due to high dust extinction (e.g., \citealt{Lada94}). Extinction maps for each of our five clouds were created using archived 2-Micron All Sky Survey (2MASS) catalogues of point sources in a similar manner described in \citet{Lada94}, \citet{Lombardi01}, \citet{Cambresy02}, and \citet{Schneider06}. For more details, see \citet{Schneider06}.

First, individual extinction values were obtained from a weighted average of the J-H and H-K colors of individual stars, assuming the average intrinsic colors were (J-H)$_0 = 0.45 \pm 0.15$ and (H-K)$_0 = 0.12 \pm 0.05$ as derived from stellar population models and typical dispersions of galactic stars measured using the Besancon Galactic model\footnote{\footnotesize http://www.obs-besancon.fr/} by  \citet{Robin03}. Second, the Besancon stellar models were used to derive average densities of foreground stars in the 2MASS bands at the distance of each cloud. For clouds near the Galactic plane or at large distances, the foreground star populations can be significant. The expected numbers of foreground stars were removed using the bluest sources (see \citealt{Cambresy02} for more details). Finally, visual extinction was derived from averaging the individual $A_V$ measurements for stars within a Gaussian beam. This averaging excluded the bluest sources and was adapted so that $\gtrsim$ 10 stars contribute to the extinction element.

\subsubsection{Regions with Complete Submillimeter - Infrared Coverage}\label{CoverageDataSection}

As an all-sky survey, 2MASS data naturally encompassed the full extent of our 5 clouds. The Spitzer coverages of these clouds were quite large, but generally restricted to areas of $A_V \ge 3$. SCUBA was used to map wide regions only rarely (see \citealt{Johnstone04}, \citealt{Hatchell05}, \citealt{Kirk06}) given its limited sensitivity and the observations typically focused on regions of known star formation.

Figure \ref{complete} compares the areal coverage from 2MASS (background) with the areal coverage from the Fundamental SLC (contours). Also included are boundaries to distinguish the Scorpius and Monoceros clouds from the Ophiuchus and Orion clouds in their respective maps. The boundaries for Serpens are not clear and much of our extinction map is associated with the Auriga system. From Figure \ref{complete}, we see that much of these clouds remain unmapped in the submillimeter, including some regions of high extinction. Table \ref{pixels} lists the cloud area observed by each of the surveys.

\subsection{Identification of Cores Within Clouds}\label{IDcoresResultsSection}

Using the Fundamental SLC we generated lists of core candidates for each of our clouds. Since our extinction maps extend over entire clouds, we used these data to determine where the core candidates were located within the large-scale structure of their parent clouds. We did this by identifying the extinction level at the position of each of our core candidates. For this study, we required that cores be located in a cloud region of $A_V \ge 3$ to ensure the cores are dense. Generally, SCUBA observed higher extinction regions, so this constraint did not greatly affect the number of core candidates in our sample. 

Also, we visually inspected our core candidates to remove those that were likely artifacts of flat-fielding or which appeared too diffuse to be dense cores. Finally, to ensure we had good detections, we removed all objects with 850 $\mu$m peak fluxes less than $5 \sigma$ (5 x 30 mJy beam\negExp\ = 150 mJy beam\negExp), where $\sigma$ is the average noise level of 30 mJy beam\negExp. Table \ref{cuts} summarizes all the cuts made to the core candidates extracted from the SLC Fundamental Catalogue. For Serpens, only one relatively small region of $\sim 1$ pc$^2$ was observed with SCUBA, resulting in far fewer core numbers relative to the other clouds.

\section{Separating Starless and Protostellar Cores: A New Classification Technique} \label{ResultsSection}

\subsection{Summary of Previous Approaches}\label{SLvsProt}

In compiling samples of cores observed in each cloud, we did not distinguish between different stages in core evolution, such as those that are starless or those that are protostellar. These populations must be separated to explore properly the relationship between the IMF and CMF (\citealt{Ward-T07GBS}). Previous efforts have attempted to separate the starless and protostellar core populations in molecular clouds by using infrared observations with (sub)millimeter continuum data. Slightly different methodologies were used, however, to accomplish this separation.  

There are many possible approaches to identifying protostellar cores. For example, both the J\o rgensen method (see J\o rgensen et al. 2006, 2007, 2008; hereafter \citetalias{Jorgensen06}, \citetalias{Jorgensen07} and \citetalias{Jorgensen08}, respectively) and the Enoch method (see Enoch et al. 2008, 2009; hereafter \citetalias{Enoch08lifetimes} and \citetalias{Enoch08young}, respectively) considered a (sub)millimeter core to be protostellar when a YSO-like infrared source was in close proximity to the dense core. These two methods, however, differed in their identification of YSO-like infrared sources and in their definition of ``close proximity''. We discuss differences between these two techniques in \S \ref{MethodComp}. Aside from these methods, other approaches included using the SEDs of point sources to identify infrared excesses, calculate spectral indices or measure bolometric temperatures (e.g., \citealt{Hatchell07}; \citealt{Winston07}; \citealt{Evans09}) and using infrared colors to identify and classify YSOs (e.g., \citealt{Allen04}; \citealt{Megeath04}; \citealt{Harvey06}; \citealt{JKirk09}; \citealt{Gutermuth08}). As a hybrid approach, \citet{Megeath09} used colors to measure SED slopes and identify protostars.

Which approach should be used? All of these methods are subject to uncertainties from unknown reddening levels and possible chance coincidences. Overall, SEDs can accurately assess whether cores are starless or protostellar (ie., through the bolometric temperature or indicators such as infrared excesses from disks), but this requires a wealth of high-resolution data at a variety of wavelengths, which is observationally expensive. In addition, resolution can vary over different wavelengths, thus complicating how objects detected at different wavelengths are associated. Combining color and spatial co-location criteria would be least biased to particular data sets such as those with high resolution or large spectral coverage. Accordingly, we describe our new core identification system in the next sections.


\subsection{New Color Criteria}\label{colorCrit}

To produce a starless CMF, one must identify and omit cores that have lost some of their surrounding envelope from outflows or from accretion onto a protostar. This task is complicated because common sources of infrared emission include galaxies and background stars along the line of sight. Fortunately, criteria have been developed for distinguishing YSOs from star forming galaxies, active galactic nuclei (AGNs), and giant stars (e.g., \citealt{Harvey06}; \citetalias{Enoch08young}; \citealt{Gutermuth08}; \citealt{Megeath09}; \citealt{JKirk09}; \citealt{Rebull09}). In general, extragalactic sources are often very faint or have unique colors, such as particular spectral signatures due to strong polycyclic aromatic hydrocarbon (PAH) emission associated with HII regions, causing an increased infrared excess at 5.8 $\mu$m and 8.0 $\mu$m. PAH emission features are not readily detected towards embedded protostars, ensuring that extragalactic sources are identified and not protostellar sources (\citealt{vanDishoeck00}; \citealt{Geers09}). Furthermore, Spitzer observations can distinguish between objects with infrared excesses from heated dust grains and from pure photospheric emission of reddened background stars. Red stellar sources, such as AGB stars, are likely to have flat spectra in the infrared, thereby distinguishing them from protostellar sources, which peak in the mid- to far-infrared (\citealt{Harvey06}). 

Young protostars still embedded in dense cores should peak in the far-infrared due to the absorption and reprocessing of light to longer wavelengths by envelopes (a characteristic envelope temperature is $T \sim 20$ K). Several recent studies have identified these sources using IRAC and MIPS colors (e.g., \citealt{Harvey06}; \citetalias{Jorgensen07}; \citealt{Megeath09}) or 2MASS colors (e.g., \citealt{Hatchell07}; \citealt{JKirk09}). Given the scatter and overlap of various objects in color or magnitude, however, there will be some objects that are not selected and conversely, not all contaminants will be removed. In addition, excesses in some bands could be the result of different physical processes. For example, protostellar outflows can interact with the surrounding molecular cloud and result in strong emission in the IRAC bands due to shocked H$_2$ gas (\citealt{Gutermuth08}). 

To identify embedded YSOs from the millions of compact infrared sources detected by Spitzer, we chose specific infrared sources based on the quality of the detection and the source colors. Since embedded protostars peak in the far-infrared, we required that our YSOs have strong detections at 24 $\mu$m and 70 $\mu$m and rising-red colors. The red colors will exclude stellar sources, which have flat colors in the infrared regime, but not extragalactic sources. Extragalactic contamination must be excluded separately. Our complete color criteria (CC) for identifying protostellar objects is listed below:  

\begin{description}
\item[CC1. ] source 24 $\mu$m or 70 $\mu$m flux has a S/N $\ge$ 5 or if the 70 $\mu$m data was poor, a S/N $\ge$ 5 in all four IRAC bands, and
\item[CC2. ] either 24 $\mu$m or 70 $\mu$m detections are real (not band-filled limits), and
\item[CC3. ] source colors are dissimilar to those of star-forming galaxies (see \citealt{Gutermuth08}), ie.,
\begin{eqnarray*}
\mbox{$[4.5]-[5.8]$}  &<& \frac{1.05}{1.2}(\mbox{$[5.8]-[8.0]$}-1), \mbox{ and}\\
\mbox{$[4.5]-[5.8]$} &<& 1.05, \mbox{ and}\\
\mbox{$[5.8]-[8.0]$} &>&1, \mbox{ and}
\end{eqnarray*}
\item[CC4a.] if detected at 24 $\mu$m, $[8.0]-[24] > 2.25$ and $[3.6]-[5.8] > -0.28([8.0]-[24])+1.88$, or
\item[CC4b.] if not detected in 24 $\mu$m, $[3.6]-[5.8] > 1.25$ and $[4.5]-[8.0]>1.4$
\end{description}

Since embedded protostars should emit strongly in the far-infrared, CC1 selects infrared sources that were well detected at 24 $\mu$m or 70 $\mu$m. MIPS, however, has a lower sensitivity than IRAC, so we may miss sources relying on strong MIPS emission. Indeed, for Orion and Taurus, there were relatively few unique 70 $\mu$m detections with respect to unique 24 $\mu$m detections. For these clouds, we modified CC1 to identify sources based on S/N $\ge$ 5 either at 24 $\mu$m or at all four IRAC bands, rather than using 70 $\mu$m. For CC2, we followed \citetalias{Enoch08young} and excluded sources that were band-filled (see \S \ref{SpitzerDataSection}) at either 24 $\mu$m or 70 $\mu$m. Since MIPS observations have a lower resolution than the IRAC bands, bandfilling at 24 $\mu$m or 70 $\mu$m could result in contamination from the wings of nearby bright sources. Such observations make the MIPS fluxes uncertain. CC3 excludes star-forming galaxies based on the prescription developed by \citet{Gutermuth08}, which selects for a growing infrared excess at 5.8 $\mu$m and 8.0 $\mu$m due to strong PAH emission. 

The colors given in CC4a and CC4b were determined using the results from the c2d survey (\citealt{Evans09}). \citet{Evans09} produced a very large sample of YSOs (1024 over 5 clouds) classified into Class 0/I, Flat, Class II, and Class III using the source SEDs. In their Figure 11, they plotted their sample in [3.6] - [5.8] and [8.0] - [24] color-color space (with different symbols for the different classes). Since the different classes clustered together into specific regions of the color-color plot, we can use the [3.6] - [5.8] and [8.0] - [24] infrared colors to identify the degree to which our infrared sources are embedded. We wish to identify all young protostars still embedded in a dusty envelope, which includes Class 0, Class I, and Flat spectral source types (\citetalias{Jorgensen08}; \citealt{Myers08}). As such, we identified the [3.6] - [5.8] and [8.0] - [24] colors from Figure 11 of \citet{Evans09} to include these spectral classes (CC4a). For objects without 24 $\mu$m detections (ie., due to the lower sensitivity), we used CC4b which adopts the [4.5] - [8.0] color criteria from \citet{Harvey07long} and the [3.6] - [5.8] color when [8.0] - [24] = 2.25 (from CC4a).

We also attempted to remove broad-line AGN contaminants using colors outlined in \citet{Gutermuth08} but found that known young protostars were frequently labeled as AGN by these criteria. For example, using this prescription and the Perseus infrared sources, the Class 0 objects HH 211, IC 348 MM, and N1333 IRAS 4B (\citealt{Froebrich05}) could be misidentified as AGN. Several authors (e.g., \citealt{Winston07}; \citealt{Gutermuth08}; \citealt{Megeath09}) apply magnitude cuts to reduce the contamination from extragalactic sources. These cuts can also remove faint emission from deeply embedded protostars. Extragalactic sources are randomly distributed in the sky. By imposing a condition that our YSO-candidates are coincident with cores (\S \ref{fluxCrit}), we expect a low probability of extragalactic contamination, and we do not apply a magnitude limit in our selection of YSO-candidates. Studies in \citetalias{Jorgensen07} and \citetalias{Jorgensen08} suggest that the likelihood of coincidences within 15$\arcsec$ of a core by a random distribution of galaxies is only $\sim 1$\%. For the most part, we will be considering larger areas than 15$\arcsec$ (see \S \ref{fluxCrit}), such that the chance of random coincidences for our data could be $\sim 4$\%, though this likelihood is probably lower since we removed sources that are likely to be be considered star-forming galaxies using CC3. 

Table \ref{myTechIRCuts} lists the number of infrared sources that met each of our color criteria. Our final source list may still include some galaxy candidates and other suspect objects, which are unlikely to be coincident with a SCUBA core. Figure \ref{colorcolor} illustrates the colors for all objects that passed CC1 and CC2. The solid lines illustrate the CC4a limits. The small light grey circles represent the infrared sources found at locations with S$_{850} < 90 $ mJy beam\negExp\ that were cut from our sample. The crosses represent the infrared sources identified as star-forming galaxies with PAH emission (CC3). The diamonds represent our final set of infrared sources. From Figure \ref{colorcolor}, it is clear that some star-forming galaxies have $[3.6]-[5.8]$ and $[8.0]-[24]$ colors that match our criteria for protostellar cores. This stresses the importance of removing extragalactic sources as well as the difficulty in identifying them.

Another complication is excluding more evolved YSOs (such as Class II sources) which are not randomly distributed but found within molecular clouds. \citet{Megeath09} calculated colors using power-law SEDs which had slopes that reflected the different classes of YSOs. For example, they find $[8.0]-[24] > 3.201$ corresponds to Flat ($-0.3 < \alpha < 0.3$) spectral sources and $[8.0]-[24] > 3.917$ corresponds to rising ($\alpha > 0.3$) spectral sources (ie., Class 0 or Class I). In comparison, we use a bluer value of $[8.0]-[24] > 2.25$, which suggests that we may be including some Class II sources in our list. We suspect, however, that this effect is minimal since most of our sources have $[8.0]-[24] > 3.201$ (see Figure \ref{colorcolor}) and most of the sources with bluer $[8.0]-[24] $ colors are very red in $[3.6]-[5.8]$.

\subsection{New Coincidence Criteria\label{fluxCrit}}

One of the main conclusions of \citetalias{Jorgensen07} was that protostars are found very close to the peaks of submillimeter emission (see their Figure 2), indicating that the motion of protostars relative to their natal cores is likely small. Therefore, it may be superior to consider a given core to be protostellar only if a YSO candidate is found relatively close to the peak intensity. It is certainly reasonable to assume, for the most part, that protostars will form in the highly concentrated parts of cores which are generally associated with the peak submillimeter flux. This could become problematic in instances when many small cores are blended together, at which point the peak value could be off-centre from all of the individual cores. 

One important consideration is that cores are typically not circular in projection, but can be very irregular in shape. Thus, a circular approximation of the core extent by using its ``effective'' radius, which can be quite large ($\sim 50\arcsec$), or a fixed angular size (ie., 15$\arcsec$ as used in the J\o rgensen method) could probe beyond the respective boundaries of the cores, such as when a core is very elongated. In using the effective radius, a larger core has a greater intrinsic chance of coincidence with a nearby infrared source.  Fixed angular sizes, however, cover different physical scales at different distances.

To ensure that the observed size and shape of the core is considered, we suggest a scheme where the object location is compared to a percentage of the \emph{difference} between the peak 850 $\mu$m intensity and the boundary intensity (90 mJy beam\negExp). Figure \ref{myTech} demonstrates such an implementation, where the search area for coincidence is given by the shaded region. We parameterized this shaded region using Equation \ref{contour}, namely; 
\begin{equation}
S_{limit} = S_{peak} - \epsilon (S_{peak} - S_{bound})\label{contour},
\end{equation}
where $S_{limit}$ is the minimum associated 850 $\mu$m flux for a YSO candidate to be considered associated with a SCUBA core, $S_{peak}$ is the peak 850 $\mu$m intensity of the core, $S_{bound}$ is the threshold intensity of the core boundary (90 mJy beam\negExp\ for the alternative flux), and $\epsilon$ is a parameter for scaling $S_{limit}$ between $S_{peak}$ and $S_{bound}$. Thus, $\epsilon$ parameterizes the proximity of a YSO-selected infrared source to the peak flux of a submillimeter core. With this coincidence criteria,  we consider both the core size and the flux distribution (core shape).

We chose a value of $\epsilon = 0.75$ after testing several different limits. Taking $\epsilon = 0.75$, while technically arbitrary, corresponds to a fractional size where the number of positive coincidences does not rise rapidly and this value is also distinct enough from the boundary (when $\epsilon = 1$) to limit the effect from uncertainties in the extent of the core. As such, we considered only cores with at least one unique YSO candidate interior to the contour for $\epsilon = 0.75$ to be protostellar. Identifying unique YSO candidates within the contour defined by $\epsilon = 0.75$ was not always clear, however. For example, some protostellar core candidates were located in filamentary structures or in crowded regions which made the surrounding SCUBA flux levels much higher. As such, there were several cases where a YSO candidate fell within the $\epsilon = 0.75$ contour of two cores. For YSO candidates associated with multiple cores, we assigned the infrared-selected source to the nearest core. 

\subsection{Method Comparison}\label{methComp}

Table \ref{compareProto} compares the numbers of protostellar cores identified using our criteria (see \S \ref{colorCrit} and \S \ref{fluxCrit}), with results from applying the J\o rgensen and Enoch core classification methods  to our SLC-derived core lists. The J\o rgensen method identified cores as protostellar if a MIPS source was found within 15$\arcsec$ (1 FWHM of the unsmoothed, non-SLC SCUBA beam) of the core centre, regardless of position angle. They also considered infrared sources of suitably red colors and cores of high central concentration, which we do not include for our implementation of the J\o rgensen method since the former is not constrained to submillimeter cores (our aim is to identify starless and protostellar cores) and the latter was shown to be a poor indicator for core evolution (see \citetalias{Jorgensen08}). Conversely, the Enoch method focused on infrared sources with certain properties (ie., certain magnitude levels, c2d class designations, and spectral indices) that were located within an intrinsic core FWHM (a geometrical mean of the core's major and minor axes), of the core centre, regardless of position angle. Since Taurus and Orion were not part of the c2d survey, we could not use the Enoch method, which selected sources with favorable c2d class designations\footnote{\footnotesize All infrared sources in the c2d catalogue are labeled according to their observed SED. Designations often associated with protostars include ``YSOc'' and ``red''. See the Final Delivery Document, Evans et al. (2007), for more details.}, however, we were able to apply the J\o rgensen method to these clouds using our modified S/N requirements;  namely, we used a S/N $\ge 5$ for at all IRAC bands in place of using a S/N $\ge 5$ at 70 $\mu$m.

We find the results from our technique are in reasonable agreement with those obtained from the J\o rgensen and Enoch methods on the same data. There are some discrepancies and any given core could be classified differently between the various methods. For example, in Perseus, we identified 46 cores as protostellar, but only $\sim 84$\% of these cores are similarly identified as protostellar with either the J\o rgensen method or the Enoch method. Since the color selection criteria and coincidence criteria between these three methods are different, some discrepancies can be expected. 

In using intrinsic core FWHMs, the Enoch method is more prone to identify cores in clustered regions as protostellar, since there are generally more YSOs in these regions and the core sizes can be quite large (thereby, considering a greater coincidence area). For both Ophiuchus and Perseus, the Enoch method classified the most objects as protostellar. In using a fixed angular scale (ie., 15$\arcsec$) the J\o rgensen method will cover a very different size scale at 125 pc (distance to Ophiuchus) compared to 450 pc (distance to Orion). As well, the SLC defined cores to have a minimum radius of 9.6$\arcsec$ (based on a minimum of eight 6$\arcsec$ pixels), therefore,  depending on the intrinsic size of a core population and the distance (and the degree in which cores are unresolved and blended), the observed size may be smaller than a 15$\arcsec$ search radius. For Taurus and Orion, the J\o rgensen method identified many more protostars than our method, which could be related to the projected core size. Indeed, the average starless core size for Taurus and Orion (28.5$\arcsec$ and 32.5$\arcsec$, respectively) are smaller than the average size for Ophiuchus and Perseus (37.3$\arcsec$ and 34.1$\arcsec$, respectively), albeit, these differences are small. For Serpens, there are not enough cores to compare well the three methods. 

In comparison, we used a size based on a fractional flux to determine our coincidence criteria. This selection method considers both core shape and size, such that a smaller intrinsic core population or more distant cloud will not be biased. In clustered regions, however, the submillimeter flux contours become complicated, so we are not fully capable of identifying these cores. As well, these clustered environments may have several bright sources that obscure or confuse the infrared emission towards nearby submillimeter cores (e.g., OMC-1, \citealt{PetersonMegeath08}). Therefore, identifying protostellar cores in clustered environments is more difficult, and we caution that extreme cases, like OMC-1, might require special consideration (see \S \ref{starlessCMFsection}). 

To eliminate the problem of chance coincidences (in both clustered and isolated environments), we applied a series of color criteria to eliminate obvious contaminants. Since we are interested in embedded sources, we select objects with strong MIPS detections (CC1 and CC2) and colors that resemble YSOs (CC4a or CC4b). In addition, we exclude galaxy candidates based on colors (CC3) rather than magnitude, since the latter will vary with cloud extinction, thereby possibly removing deeply embedded protostars. Our criteria made no initial assumptions regarding c2d designation, and yet most of our protostellar sources in Ophiuchus, Perseus, and Serpens have c2d designations often associated with YSOs, suggesting that our method is capable of identifying protostellar sources without initially relying on c2d designations. This consistency allows us to apply our technique to data observed differently from the c2d clouds (Taurus and Orion) that have no such designations. There are, however, two protostellar cores with c2d designations of ``galaxy candidates'' in Perseus and one each in Ophiuchus and Serpens.

It is important to note that the J\o rgensen method and the Enoch method were developed based on different data sets. The J\o rgensen method employed non-SLC SCUBA maps with a pixel resolution of 3$\arcsec$ and a beam angular resolution $\sim 15\arcsec$, whereas the SLC has a pixel resolution of 6$\arcsec$ and a smoothed beam angular resolution $\sim 23\arcsec$ at 850 $\mu$m. Thus, the 15$\arcsec$ distance limit used in the J\o rgensen method may be inaccurate for the SLC data. The Enoch method was developed using cores identified at 1.1 mm (not 850 $\mu$m) with Bolocam ($40\arcsec$ FWHM resolution) for their respective analyses, and so the core numbers, locations and sizes from each sample will differ. 

\subsection{Comparison to Known Class 0 Sources}

\citet{Froebrich05} compiled a database of young protostars from the literature, and used all available photometry between 1 $\mu$m and 3.5 mm to build SEDs and re-classify the sources under a consistent system. This sample includes two objects in Ophiuchus and Taurus, ten objects in Perseus, and eighteen in Orion. For the most part, we identified objects in common as protostellar. Some protostars listed by \citet{Froebrich05}, however, have no corresponding submillimeter cores in the Fundamental SLC. For example, the known protostar VLA 1623, which is located $\sim 0.5\arcmin$ southwest of the bright and very crowded Oph A filament in Ophiuchus, does not have a submillimeter core in the Fundamental SLC. This core is listed in the Extended SLC, which we do not consider (see \S \ref{ScubaDataSection}).

In addition, some known young protostars were misidentified as starless cores due to no or poor infrared detections with Spitzer. For example, two protostars in Perseus, SVS 13B and NGC 1333 IRAS 4A, had negative bandfilled IRAC fluxes and thus, failed our color criteria. For the most part, there is good agreement between our classification and the compilation from \citet{Froebrich05}, thus our technique is capable of identifying protostellar cores, provided that Spitzer was able to detect the source. Indeed, we are limited in our ability to classify our SCUBA cores by the areal coverage of Spitzer in relation to the SCUBA observations. To be as unbiased as possible, we did not alter our lists to include known protostellar cores that were labeled starless with our technique. We will consider separately the samples of starless and protostellar cores in the next section. 


\section{Discussion}\label{DiscussionSection}
One of our main goals is to examine CMFs for a variety of nearby clouds. In the previous section, we applied our own classification technique to identify which cores were starless and which were protostellar. To produce the CMFs, we must first estimate the mass contained within the cores. The mass of a SCUBA object can be estimated from the Planck function, 850 $\mu$m flux, opacity, and distance. For example, using the formula (\citealt{Johnstone00}):
\begin{equation}\label{massEq}
M_{clump} = 0.19\ S_{850}\left[\exp{\left(\frac{17\ K}{T_d}\right)} - 1 \right] \left(\frac{\kappa_{850}}{0.01\ \mbox{cm}^2\ \mbox{g}^{-1}}\right)^{-1} \left(\frac{D}{160 \mbox{ pc}}\right)^2 M_{\odot}
\end{equation}
where $S_{850}$ is measured in Jy. For all clouds, we assumed a constant dust opacity, $\kappa_{850} = 0.01$ cm$^2$ g\negExp, though $\kappa_{850}$ can vary by a factor of two (\citealt{Henning95}). Our cloud distances and dust temperatures are listed in Tables \ref{cloudMass} and \ref{prop}, respectively. We also assumed a single uniform temperature for all cores in a given cloud. Temperature distributions across cloud cores are not yet well known. Note that dust temperatures in starless cores should decrease from core edge to core centre as external heating from the interstellar radiation field is increasingly damped (\citealt{difran07}). Protostellar cores, however, may have internal heating which will affect the radial temperature variations, such as a warm centre, cool middle and warm edges.  

Tables \ref{starlessList} and \ref{protoList} in Appendix A list the starless and protostellar cores in Perseus identified using our classification technique. Each table lists core names (positions), peak fluxes, total fluxes, masses, and effective radii. The quoted total fluxes and effective radii are the ``alternative flux'' and ``alternative radius'', respectively (see \S \ref{ScubaDataSection}). For complete tables of all 5 clouds, see the electronic edition.

\subsection{Starless CMFs} \label{starlessCMFsection}

Figure \ref{CMF+IMF} shows the observed starless CMFs for the Ophiuchus, Taurus, Perseus, Serpens, and Orion clouds. These starless cores were identified using our classification scheme (see \S \ref{colorCrit} and \S \ref{fluxCrit}). To estimate errors in each mass bin, we varied the temperature in steps of $\pm 1$ K up to a 30\% uncertainty in $T_d$, and found the standard deviation across the mass bins generated from the CMFs at these temperatures. We found such $\Delta T$ standard deviations were similar in magnitude to errors expected from Poisson statistics. In each panel of Figure \ref{CMF+IMF}, a power-law relationship with a Salpeter slope ($-1.35$, see \S \ref{IMFrelation}) is also shown as a dotted line and this seems to trace reasonably well the higher mass ends of the CMFs.

Since we use the same classification scheme for all five clouds, differences in the CMFs should reflect differences between core populations in the clouds. For example, Orion contains more massive starless cores than Ophiuchus or Taurus, though many of the highest mass cores in Orion could be misidentifications associated with OMC-1 (see below). Conversely, Ophiuchus and Taurus contain some very low-mass cores. This could be related to cloud distance (see Table \ref{cloudMass}), as Ophiuchus (125 pc) and Taurus (140 pc) are much closer than the other clouds, but a similar difference is not seen between Perseus (250 pc) and Orion (450 pc).

In general (but excluding Serpens), the CMFs have a similar ``lognormal'' appearance, but some differences between them remain. Perseus shows the narrowest distribution. Orion has the widest distribution, due to the extreme high-mass extent.  These high mass ``cores'' have masses on order of $\sim 10^{2-3}$ \Msun\ and may be larger-scale structures than cores. Recall that we have defined a core as a dense, compact structure that could form a single star or a small stellar system. For Ophiuchus, Taurus, Perseus, and Orion, the CMF peaks are at roughly 0.1 \Msun, 0.2 \Msun, 1.6 \Msun, and 1.0 \Msun, respectively. The Serpens CMF has a very small sample, and we cannot make the same conclusions or comparisons. 

As discussed before in \S \ref{methComp}, the clustered region, OMC-1, in Orion is a very complex region. OMC-1 contains bright infrared nebulosity which lowers the sensitivity to protostars and could result in cores misidentified as starless. In addition, OMC-1 is exposed to intense UV radiation due to nearby OB associations (\citealt{PetersonMegeath08}). Therefore, it is also possible that we are under-estimating the temperature (20 K) in OMC-1. Indeed, increasing the temperature of these cores by a factor of 2 reduces the core masses by a factor of 2.5. To understand the impact the OMC-1 cores have on the Orion CMF, Figure \ref{CMF-OMC1} shows the Orion starless CMF after the cores specifically associated with OMC-1 (12 starless cores) are removed. Excluding these objects effectively removed the highest mass cores ($M > 15$ \Msun)  and made the CMF thinner, suggesting that the cores in OMC-1 may be different from the rest of Orion. To remain unbiased to our classification technique, however, we will mainly consider the total Orion CMF for the following analysis. We will also briefly compare the differences with the OMC-1 cores removed for the discussion of starless CMFs only. 

It is important to note that SCUBA (with a limited detection sensitivity) and Clumpfind (using a threshold of 90 mJy beam\negExp) are more sensitive to concentrated, bright cores. As such, our CMFs are incomplete for cores of low surface brightness. More sensitive instruments (i.e., SCUBA-2) will locate better these low surface brightness objects in the future, but we are unable to include these objects at present. For the most part, low surface brightness cores are expected to occupy the low-mass regime of the CMF, though there can be cases where high-mass ``cores'' are too diffuse to detect. The high-mass ends of our CMFs have been comparatively well sampled and these data can be compared best to the IMF.

\subsubsection{Relation to the IMF}\label{IMFrelation}

Using an observed luminosity function for nearby stars, \citet{Salpeter55} found that the stellar mass distribution (or IMF) seemed to obey a power-law relation, $\xi(m) \propto m^{-1.35}$ for $0.4$ M$_{\odot} \lesssim $ M $\lesssim 10$ \Msun.
Here, $\xi(m)$ is defined as $dN = \xi(m) d\log{m}$, where $dN$ is the number of stars of mass $m$ lying between $m$ and $m+dm$ (\citealt{Warner61}). Thus,
\begin{eqnarray}
\frac{dN}{d\log{m}} &\propto& m^{-1.35}  \mbox{\hspace{6mm} or}\label{IMFsal}\\[2mm]
\frac{dN}{dm} &\propto& m^{-2.35}.
\end{eqnarray}
In Figure \ref{CMF+IMF}, the Salpeter power-law slope appears to agree with the cloud CMFs at higher masses. To test this relationship more quantitatively, we calculated the ordinary least squares slope to the high-mass end of each CMF weighting each bin by its uncertainties. Since quantifying a high-mass end is uncertain, we determined several best fit slopes using different mass ranges with at least 4 mass bins to calculate the best fit slope. Figure \ref{bestFit} shows the resulting best fits (solid lines) for one choice of mass range (dotted lines). For Orion, there is a slight increase in cores at very high masses ($\sim 10^{2-3}$ \Msun) mainly associated with OMC-1 (see Figure \ref{CMF-OMC1}) and since these objects may not be cores, we do not consider these massive objects in our best fit slopes in Orion. We also calculated the best fit slopes for the Orion CMF with the cores towards OMC-1 removed using similar mass ranges. Table \ref{bestFitResults} gives the average ordinary least squares best fit for the well-sampled clouds and the best fit for Orion without the 12 starless cores from OMC-1. The Serpens CMF had too few mass bins to calculate a best fit slope. 

The slopes in Table \ref{bestFitResults} hint at some differences, though all agree with the Salpeter power-law slope, within $\sim 2 \sigma$ errors.  Indeed, some clouds have CMFs that are more Salpeter-like than other clouds. The errors were derived from the best fit and reflect the shape of the CMFs. For example, a CMF that appears curved (e.g., see Orion) will have a larger uncertainty associated with a linear slope than a CMF that follows a more linear drop-off (e.g., see Taurus). For the Orion CMF without the OMC-1 cores (see Figure \ref{CMF-OMC1}), the shape is less curved and the errors more similar to what is found for Ophiuchus and Perseus (see also, Table \ref{bestFitResults}). The slope obtained for the complete Orion CMF  ($-1.69\pm 0.72$) flattens considerably when the cores from OMC-1 are removed ($-0.93 \pm 0.18$). 

In addition to producing ordinary least squares fits, we also used the Kolmogorov-Smirnov (KS) Test on the high-mass end of our CMFs. We generated random core masses following a Salpeter distribution within the mass ranges listed in Table \ref{bestFitResults} and determined the likelihood that our observed CMFs were drawn from the same distribution as the random Salpeter sample. Unfortunately, the likelihood varied significantly with mass range, suggesting that our samples are too small for the KS test. A larger, more sensitive set of observations would significantly improve the comparison with the Salpeter IMF.  

\subsubsection{Trends with the CMFs}\label{trendsSection}

Our five clouds represent a variety of environments and have very different properties (see \S \ref{CloudsSection}). Any trends between these properties with the starless CMFs could reveal information about star formation across these environments. We compared the CMF peak mass and the best fit slope against cloud distance, cloud mass, cloud extinction peak, core line widths, core temperature, core extinction peak, and number of cores in the sample. We found very few correlations, in spite of our range of cloud properties. Figure \ref{trends} shows four of the fourteen trends that we examined.  

Figure \ref{trends}a shows no correlation between core temperature and peak mass, whereas Figure \ref{trends}b (core temperature with best fit slope) shows the strongest case for a potential trend. The correlation in Figure \ref{trends}b, however, depends greatly on the slope of Orion, which has the largest measured uncertainty. If you consider the slope obtained for Orion excluding the cores towards OMC-1 ($-0.93 \pm 0.18$), this trend disappears. A trend between core temperature and best fit slope would imply that there is no universal power-law distribution for CMFs, unless starless cores exist at the same temperature. A more extensive study with more clouds is necessary to test this relationship (note, that our core temperatures are not particularly well-defined - see \S \ref{coreProp}). Figure \ref{trends}c shows a possible trend between cloud distance and CMF peak mass. A slight positive relation between these properties may reflect our imposed threshold of 90 mJy beam\negExp. From Equation \ref{massEq}, mass should vary with distance and temperature, however there is no strong correlation between peak mass and temperature (Figure \ref{trends}a). The weak trend in Figure \ref{trends}c could also result from resolution problems, such as numerous lower mass cores appearing as a single higher mass cores in low resolution distant clouds. This would suggest, however, a shallower best fit slope with distance, which we do not observe in our furthest cloud, Orion, unless we exclude the OMC-1 cores. Figure \ref{trends}d shows no correlation between best fit slope and the number of cores.

\subsubsection{CMFs Using Different Classification Methods}\label{MethodComp}

We applied two additional techniques in classifying a core as protostellar or starless: the J\o rgensen method and the Enoch method. The J\o rgensen method is outlined in \citetalias{Jorgensen06}, \citetalias{Jorgensen07}, and \citetalias{Jorgensen08}, and used Spitzer c2d data and non-SLC SCUBA observations for Perseus (\citetalias{Jorgensen06}) and Ophiuchus (\citetalias{Jorgensen07}), whereas the Enoch method, outlined in \citetalias{Enoch08lifetimes} and \citetalias{Enoch08young}, used Spitzer c2d data and Bolocam 1.1 mm observations for Ophiuchus, Perseus, and Serpens.

Figures \ref{CMFcompareOph} and \ref{CMFcomparePer} compare our starless CMFs for Ophiuchus and Perseus, respectively, with the starless CMFs obtained from classifying cores with the J\o rgensen and Enoch methods. Although there are relatively few cores in general with large masses, the high-mass end of the CMF appears to be vary slightly with the classification technique. Table \ref{compareHighMass} compares the number of starless high mass cores in Ophiuchus and Perseus using the three techniques. We considered one mass range from our best fit results (see Table \ref{bestFitResults}) to define the high-mass regime in each case. The Enoch method tends to classify the most massive cores as protostellar, resulting in the fewest that are identified as starless. Indeed, a similar result by \citetalias{Enoch08lifetimes} led them to conclude that higher mass cores evolve more quickly than lower mass cores. All three methods, however, still produce CMFs that agree with the Salpeter power-law distribution (dotted lines in Figures \ref{CMFcompareOph} and \ref{CMFcomparePer}). The low-mass ends also show some variations (especially in Ophiuchus), but due to incompleteness in this regime, we cannot determine the significance of these differences.

\subsection{Protostellar CMFs}\label{ProtoResults}

In identifying starless cores with our classification technique, we also identify the protostellar core population. Figure \ref{protoCMF} shows the protostellar CMFs for each cloud and their relation to a Salpeter-like slope. The protostellar cores in Figure \ref{protoCMF} were identified using our classification technique. We initially assumed the same temperatures as with the starless CMFs and uncertainties were measured in the same manner as the starless CMFs. 

The protostellar CMFs do not show quite the same ``lognormal'' shape as the starless CMFs, though there are fewer numbers of protostellar cores than starless cores. Also, all protostellar CMFs seem to be systematically shifted to slightly higher masses than the starless CMFs. The Ophiuchus and Taurus protostellar CMFs peak at higher masses  ($\sim 0.5$ \Msun\ for both) than their starless distributions ($\sim 0.2$ \Msun\ for both). The Perseus protostellar CMF peaks at $\sim 2.5$ \Msun\ and the Orion protostellar CMF peaks at $\sim 6.3$ \Msun, which are both higher than the peak for the starless CMFs ($\sim 1.0$ \Msun).  

Figure \ref{protoCMF} seems to suggest that protostellar cores are intrinsically more massive than starless cores, but we assumed they have identical dust temperatures to the starless cores. Protostellar cores, however, may be heated internally and thus, have slightly higher temperatures (e.g., see \citetalias{Enoch08lifetimes}). A higher protostellar core temperature will reduce the core mass and possibly remove the apparent discrepancy between the starless and protostellar CMFs. 

We tested the effect of higher temperatures on the protostellar CMFs by raising the protostellar core temperatures in steps of $\Delta T_d = 1$ K. We determined our final protostellar core temperature based on fitting the width and peak of the protostellar CMF to the starless CMF of a given cloud, assuming that these widths and peaks are similar and any envelope mass loss around a protostar is insignificant. In the end, we found good fit temperatures for the protostellar cores to be 20 $\pm\ 2$ K for Ophiuchus, 17 $\pm\ 2$ for Taurus, $15 \pm\ 1$ K  for Perseus, and 59 $\pm\ 3$ K for the entire Orion CMFs (Serpens has too few cores to properly match the width and peak of the CMF distributions). Thus, the protostellar core temperatures in Ophiuchus, Taurus, and Perseus should increase by a factor of $\sim 1.3$ and the temperatures in Orion should increase by a factor of $\sim 3$. Figure \ref{CMFcompare} illustrates the protostellar CMFs at their new temperatures in relation to the starless CMFs. The protostellar CMFs in Figure \ref{CMFcompare} show no overabundance in protostellar cores at higher masses relative to starless cores at the new temperatures, suggesting that higher mass cores do not necessarily evolve more quickly than lower mass cores, assuming insignificant mass loss.

The Orion protostellar CMF has a substantial temperature increase (59 K versus 20 K) compared to the other clouds, which may be related to  ``hot'' cores often associated with massive star forming regions (\citealt{Kurtz00}). Hot cores are very dense and can have temperatures of $> 50$ K (\citealt{Cesaroni94}). Figure \ref{CMF-OMC1} shows the Orion starless CMF with the high-mass star forming OMC-1 region removed. The resulting starless CMF is considerably more ``lognormal'' and consistent with the other clouds. This suggests that the cores in OMC-1 may be of unusually high-mass or at a higher temperature. 

For the rest of our analysis, we will compare starless and protostellar core properties using the same dust temperatures (see Table \ref{prop}) and not the increased temperatures shown in Figure \ref{CMFcompare}. In addition, for the rest of our analysis, we will consider the Orion core population with OMC-1 as opposed to removing the 12 starless cores in OMC-1 from our sample.

\subsection{Core Environments} \label{coreEnviro}

Figure \ref{BEmodels} compares the masses and the observed radii of the cores in our five clouds. The observed radii are the true radii convolved with the SLC beam and then truncated according to the 90 mJy beam\negExp\ contour threshold. The SLC beam is approximated as a Gaussian with a FWHM of $22.9\arcsec$ (\citealt{difran08}). Due to this finite resolution, all cores will have smaller true sizes than observed, with less extended cores more affected than larger ones. We have not deconvolved the sizes of our cores in Figure \ref{BEmodels}. 

For all the clouds in our sample, core mass and radius are well correlated. Figure \ref{BEmodels} also includes the sensitivity limit (dot-dashed line) from the 90 mJy beam\negExp\ threshold. Although the scatter is quite restricted in one direction due to sensitivity, we note that at higher masses, core mass begins to rise faster than size in Figure \ref{BEmodels}. Indeed, Figure \ref{BEmodels} illustrates our completeness level. At the lowest masses, we are very close to the limit of our ability to detect cores (closer to our sensitivity limit). For Serpens, the protostellar cores appear to have higher masses and larger sizes than the starless cores. This trend, however, may be a reflection of the small sample, since the other clouds do not show the same results. 

Figure \ref{AV-dist} gives the $A_V$ distributions for the five clouds, separated by protostellar (solid lines) and starless (dashed lines) cores. The histograms are all binned to $\Delta A_V = 4$ to ensure each bin is well populated. Within a given cloud, the distributions are broadly similar. For example, both the protostellar and starless core $A_V$ distributions for Ophiuchus peak at $A_V \sim 25$ magnitudes. This peak is at significantly higher $A_V$ than the peaks in the other clouds, though this could be related to distance. For more distant clouds, we are likely less sensitive to higher extinctions due to lower linear resolution. Serpens is the only exception to the similar peaks in $A_V$ between protostellar and starless core populations, though that is likely due to its small sample size. 

In Figures \ref{AV-Mass} and \ref{AV-Rad} we compare extinction with core mass and core size, respectively. For Ophiuchus, Taurus, Perseus, and Orion, we found no strong trends between these distributions. Serpens may show trends of increasing mass and size with extinction for starless to protostellar cores, though the other clouds with larger samples do not reveal the same trends. For the other clouds, the protostellar and starless cores have wide ranges of mass and size over a wide range of extinctions. This finding is counter to the expectation that cores at higher extinction could be at higher pressures and hence may be smaller in size (\citealt{Johnstone04}). While there is a correlation between core size and mass (see Figure \ref{BEmodels}) for all five clouds, no obvious relations of these quantities with extinction are seen given a good sample size.

\subsection{Predicted CMFs}\label{predCMFSection}

SCUBA was limited in its ability to sample the dense core populations in nearby star-forming clouds and typically observed known regions of active star formation (e.g., L1688, \citealt{Johnstone00}), so very large fractions of these clouds, particularly at low extinctions, remain unmapped (see Figure \ref{complete}). As a result, the CMFs derived from SLC map data (see \S \ref{starlessCMFsection}) do not represent all cores within their respective clouds.

SCUBA-2 will be the successor continuum mapping instrument to SCUBA on the JCMT. The SCUBA-2 Gould Belt Legacy Survey (GBLS) will map 15 star-forming molecular clouds within 500 pc with excellent sensitivity to detect core down to substellar masses (\citealt{Ward-T07GBS}). These data will allow us to obtain robust CMFs, construct a less ambiguous approach to classifying cores, constrain the lifetimes associated with star formation, and solidify any possible similarities between CMFs and the IMF. With superior sensitivity and efficiency, SCUBA-2 will be able to map nearby clouds with improved flux sensitivity and speed, allowing for more complete samples of their core populations.

We wish to extrapolate our current CMFs to predict the CMFs we will obtain using SCUBA-2. Predicted CMFs can be created by assuming the observed CMFs trace the incidences of cores for the unobserved regions of each cloud. Of course, since SCUBA-2 will have improved sensitivity, lower surface brightness cores will be more easily detected. As well, our sample is limited to cores with minimum peak fluxes of 150 mJy beam\negExp\ and size limits of 90 mJy beam\negExp\ (following our original criteria for identifying cores; see \S \ref{IDcoresResultsSection}). Thus, we caution that our predicted CMFs are only applicable in the higher mass regime where our observed data are more complete.

The predicted starless CMFs will be extrapolations of the observed starless CMFs over the unobserved regions of each molecular cloud. We considered two techniques to create our predicted CMFs. First, an $A_V$-independent extrapolation where we gave each core equal weight and extrapolated the observed starless CMFs using the ratio of total cloud area (for $A_V > 3$)  to cloud area observed by SCUBA. Second, an $A_V$-dependent extrapolation, where we extrapolated core incidences over small extinction ranges using the ratio of total cloud area over ranges in $A_V$ to cloud area observed by SCUBA over the same ranges of $A_V$. Equation \ref{predEq} outlines the $A_V$-dependent extrapolation. We chose $\Delta A_V = 4$ to ensure that the extinction ranges are well populated. For a given $A_V$ range, the predicted number of objects ($N_{A_V, pred}$) is given by the observed number in that range ($N_{A_V, obs} $) multiplied by the ratio of entire cloud area to the observed cloud area at that $A_V$ range.
\begin{equation}\label{predEq}
N_{A_V, pred} = N_{A_V, obs} \frac{\Omega_{A_V, cloud}}{\Omega_{A_V, scuba}}
\end{equation}
where $\Omega_{A_V, cloud}$ is the area of the whole given cloud that falls within a given $A_V$ range and $\Omega_{A_V, scuba}$ is the area within the same extinction range observed by SCUBA. 


The fraction of each cloud observed by SCUBA was different from cloud to cloud. Thus, taking into account the area of observations can help us compare the results between clouds. Each SLC map is $1.2\arcmin$ x $1.2\arcmin$ in extent and several overlap with other maps (\citealt{difran08}). Table \ref{pixels} lists the total areas observed by SCUBA for each cloud. 

One additional complication is that the Scorpius and Monoceros clouds have projected locations near Ophiuchus and Orion, respectively (see Figure \ref{complete}). As such, the 2MASS extinction maps of Ophiuchus and Orion contain sections of Upper Scorpius and Monoceros R2. For Serpens, we consider the entire 2MASS map since the cloud boundary is not clear as with the other clouds. Since our core counts with Serpens are quite low, this will not greatly affect our overall results.

\subsubsection{Predictions}\label{predVals}

The predicted CMFs are shown in Figure \ref{predCMFs+obsCMFs} for each cloud, comparing the observed starless CMF (dotted histogram) to the $A_V$-dependent extrapolated (solid histogram) and $A_V$-independent extrapolated (dashed histogram) starless CMFs. Overall, the predicted starless CMFs are generally similar in shape, even in the $A_V$-dependent case, but have large differences in number. 

Table \ref{predCoreNum} lists the numbers of observed starless cores and the numbers of predicted starless cores using the above assumptions. The quoted errors in our predicted core counts were determined from a quadrature sum of the $\sqrt{N}$ uncertainties in each bin of the observed CMF scaled upwards according to the ratio of unobserved to observed area. The errors given in Table \ref{predCoreNum} show that the difference between the predicted numbers and the observed numbers are significant. Our analysis, however, is biased against low-mass objects, particularly diffuse sources, which were more difficult to detect or keep within our core definition limits. Since we cannot accurately probe the low-mass end of the observed CMFs, our predicted CMFs are also incomplete in this regime and the numbers quoted in Table \ref{predCoreNum} are lower limits to the actual expected core statistics using these methods. Nevertheless, we predict increases in the total numbers of starless cores by factors of at least 6 - 11 in future surveys.

\setlength{\hdashlinewidth}{.8pt}
\setlength{\hdashlinegap}{2pt}

There is a significant increase in the number of predicted cores obtained from using the $A_V$-independent extrapolation over the numbers from the $A_V$-dependent case. Since the $A_V$-independent method predicts so many more cores over the $A_V$-dependent method and the majority of the cloud exists at lower extinction levels, this increase suggests that cores observed by SCUBA tended to be located in regions of relatively high extinction. Indeed, SCUBA mainly focused on the very high extinction regions of each cloud (see Figure \ref{complete}).

\subsubsection{Gaussian Replacements}\label{gausRep}

The predicted CMFs in Figure \ref{predCMFs+obsCMFs} were made assuming we accurately knew the mass of each core based on flux and temperature. This assumption, however, may be unwarranted since uncertainties from the instruments, temperature gradients, distance measurements, and cloud opacities will all affect the estimated core mass. We attempted to simulate this uncertainty by replacing core populations in every mass bin of the observed CMFs with a Gaussian of the same area. We set the FWHMs of our Gaussian replacements to be the width of our initial mass bins ($\log{M/M_{\odot}} = 0.2$) and with these replacements, we computed predicted CMFs with the same assumptions as before.

Figure \ref{predCMFs+fit+gaus} shows the predicted starless CMFs assuming an $A_V$-dependent extrapolation and Figure \ref{predPixCMFs+fit+gaus} shows the predicted starless CMFs for the $A_V$-independent extrapolation. Included in both figures are the cumulative Gaussian replacements for each of the mass bins and a Salpeter power-law slope for comparison. The Gaussian replacements show a low-mass ``tail'' due to projecting the Gaussians from a linear plane to a logarithmic plane. The errors in the bin populations were determined by varying the temperature with the observed starless CMFs as before. The slopes of the predicted histogram distributions and the predicted Gaussian replacement distributions all agree with the Salpeter power-law, within errors. 

\section{Conclusions}\label{conc}

We have used data from SCUBA, Spitzer, and 2MASS to produce relatively unbiased starless and protostellar CMFs using a new classification technique. We applied this technique to the Ophiuchus, Taurus, Perseus, Serpens, and Orion molecular clouds. To our knowledge, this work is the most extensive study of the starless core populations in nearby molecular clouds to date, since we have examined 5 unique clouds in a uniform manner where previous studies have focused on smaller regions or fewer clouds (e.g., \citetalias{Jorgensen07}; \citealt{Nutter07}; \citetalias{Enoch08lifetimes}; \citealt{Nutter08}) with different core identification schemes.

It is important to consider that our core populations are incomplete. We were unable to detect well cores of low surface brightness (diffuse) with SCUBA. These cores will be more easily observed with more sensitive instruments, such as SCUBA-2. We expect that the low-mass end of our CMFs will be most affected by the improved sensitivity of SCUBA-2. The high-mass end, however, should be relatively complete. As well, we have developed a self-consistent technique to analyze these five clouds and determine which cores were starless and which were protostellar. Our technique identifies YSO-like infrared sources through strong infrared emission and red colors associated with embedded protostars. We then used the 850 $\mu$m flux at the position of each of these YSO-like sources to determine which were associated with our submillimeter cores. With our classification technique, we developed uniform samples of cores for each cloud.

Assuming a constant dust temperature for all cores in a given cloud, our starless CMF best fit slopes were $-1.26 \pm 0.20$, $-1.22 \pm 0.06$, $-0.95 \pm 0.20$, and $-1.67 \pm 0.72$ for the Ophiuchus, Taurus, Perseus, and Orion starless CMFs, respectively. Removing the OMC-1 cores from Orion flattened the slope to $-0.93 \pm 0.18$. These slopes all agree with the Salpeter power-law (Salpeter $= -1.35$) within errors, though it is possible that the OMC-1 region represents a population of cores with very different properties with respect to the rest of the Orion cloud. We were unable to calculate a best fit slope for Serpens due to the small sample size. 

Our starless CMFs showed similar log-normal shapes, though the range of masses and peak in the distribution varied with each cloud. The starless CMF for Orion also included an unusual population increase at very high masses due to cores in OMC-1, which could be attributed to its cores having a more varied temperature or to difficulties in distinguishing between starless and protostellar cores with the infrared data. Nevertheless, with our self-consistent technique, our four well sampled clouds all agree with a Salpeter power-law distribution within $2 \sigma$ errors (except Orion without OMC-1 which is $\sim 2.3 \sigma$). To better understand the shape of these CMFs and their relation to the IMF, we need larger samples to minimize the uncertainties. 

We also tested our classification scheme against two others (the J\o rgensen and Enoch methods) and found that the fraction of high-mass cores identified as starless or protostellar varied slightly with the technique used for classification. In addition, we also examined trends in the starless CMF peak masses or best fit slopes with cloud properties, and we found a potential relationship between best fit slope and core temperature, suggesting that a single power-law relationship for CMFs may not be ideal. This correlation, however, depends greatly on the slope of the Orion CMF, which has the largest uncertainty. Indeed, this trend disappears if we consider the Orion CMF with the cores towards OMC-1 removed. Since the CMFs all agree with a Salpeter distribution within errors, more data at better resolution and good sensitivity are necessary to understand these relations.

Assuming the same dust temperature as the starless cores, the protostellar CMFs were systematically shifted to slightly higher masses. We increased the protostellar core temperatures to fit the width and peak of the protostellar CMFs with their starless CMFs and found the protostellar core temperatures should increase by a factor of $\sim 1.3$ in Ophiuchus, Taurus, and Perseus, and a factor of $\sim 3$ in Orion over our original dust temperatures. Though these increases are plausible, radiative transfer modeling of protostellar cores is necessary to determine if protostellar heating has such a significant impact on the average core temperature. 

Finally, we used simple assumptions to predict the core populations for the unobserved regions of each cloud. We developed two sets of predicted CMFs, an $A_V$-dependent extrapolation and an $A_V$-independent extrapolation, both which agreed with a Salpeter power-law distribution. We found that the $A_V$-independent extrapolation predicted a larger number of dense cores in each cloud than the $A_V$-dependent extrapolation had predicted. We hope to compare our predicted number of cores to the results from surveys of entire cloud regions in the future (ie., with SCUBA-2).

\vspace{1cm}
\acknowledgments{This work was possible with funding from a Natural Sciences and Engineering Research Council of Canada CGS award and a Discovery Grant. The authors thank the anonymous referee for their comments and suggestions towards this paper. The authors also thank the Taurus Spitzer Team and the Orion Spitzer Team for contributing infrared data prior to publication. SIS would like to thank J. J\o rgensen and M. Enoch for their kind time and attention to various inquiries. As well, the authors thank E. Ledwosinska, T. MacKenzie, H. Kirk, and D. Johnstone for their work in creating the SLC and N. Evans II, P. Harvey, M. Dunham, T. Huard, T. Brooke, M. Enoch, N. Chapman, L. Cieza, and K. Stapelfeldt for their work in creating the c2d catalogue.}
\acknowledgments{The James Clerk Maxwell Telescope is operated by The Joint Astronomy Centre on behalf of the Science and Technology Facilities Council of the United Kingdom, the Netherlands Organisation for Scientific Research, and the National Research Council of Canada. This work is based [in part] on observations made with the Spitzer Space Telescope, which is operated by the Jet Propulsion Laboratory, California Institute of Technology under a contract with NASA. This publication makes use of data products from the Two Micron All Sky Survey, which is a joint project of the University of Massachusetts and the Infrared Processing and Analysis Center/California Institute of Technology, funded by the National Aeronautics and Space Administration and the National Science Foundation.

\bibliographystyle{apj}
\bibliography{references}

\begin{thebibliography}{69}
\expandafter\ifx\csname natexlab\endcsname\relax\def\natexlab#1{#1}\fi

\bibitem[{{Allen} {et~al.}(2004){Allen}, {Calvet}, {D'Alessio}, {Merin},
  {Hartmann}, {Megeath}, {Gutermuth}, {Muzerolle}, {Pipher}, {Myers}, \&
  {Fazio}}]{Allen04}
{Allen}, L.~E., {Calvet}, N., {D'Alessio}, P., {Merin}, B., {Hartmann}, L.,
  {Megeath}, S.~T., {Gutermuth}, R.~A., {Muzerolle}, J., {Pipher}, J.~L.,
  {Myers}, P.~C., \& {Fazio}, G.~G. 2004, \apjs, 154, 363

\bibitem[{{Andr{\'e}} \& {Saraceno}(2005)}]{AndreSaraceno05}
{Andr{\'e}}, P., \& {Saraceno}, P. 2005, in ESA Special Publication, Vol. 577,
  ESA Special Publication, ed. A.~{Wilson}, 179--184

\bibitem[{{Andr\'{e}} {et~al.}(2000){Andr\'{e}}, {Ward-Thompson}, \&
  {Barsony}}]{Andre00}
{Andr\'{e}}, P., {Ward-Thompson}, D., \& {Barsony}, M. 2000, Protostars and
  Planets IV, 59

\bibitem[{{Bonnell} {et~al.}(2004){Bonnell}, {Vine}, \& {Bate}}]{Bonnell04}
{Bonnell}, I.~A., {Vine}, S.~G., \& {Bate}, M.~R. 2004, \mnras, 349, 735

\bibitem[{{Cambr{\'e}sy} {et~al.}(2002){Cambr{\'e}sy}, {Beichman}, {Jarrett},
  \& {Cutri}}]{Cambresy02}
{Cambr{\'e}sy}, L., {Beichman}, C.~A., {Jarrett}, T.~H., \& {Cutri}, R.~M.
  2002, \aj, 123, 2559

\bibitem[{{Cesaroni} {et~al.}(1994){Cesaroni}, {Churchwell}, {Hofner},
  {Walmsley}, \& {Kurtz}}]{Cesaroni94}
{Cesaroni}, R., {Churchwell}, E., {Hofner}, P., {Walmsley}, C.~M., \& {Kurtz},
  S. 1994, \aap, 288, 903

\bibitem[{{Curtis} \& {Richer}(2009)}]{CurtisRicher09}
{Curtis}, E.~I., \& {Richer}, J.~S. 2009, ArXiv astro-ph/0910.4070

\bibitem[{{Di Francesco} {et~al.}(2007){Di Francesco}, {Evans}, {Caselli},
  {Myers}, {Shirley}, {Aikawa}, \& {Tafalla}}]{difran07}
{Di Francesco}, J., {Evans}, II, N.~J., {Caselli}, P., {Myers}, P.~C.,
  {Shirley}, Y., {Aikawa}, Y., \& {Tafalla}, M. 2007, in Protostars and Planets
  V, ed. B.~{Reipurth}, D.~{Jewitt}, \& K.~{Keil}, 17--32

\bibitem[{{Di Francesco} {et~al.}(2008){Di Francesco}, {Johnstone}, {Kirk},
  {MacKenzie}, \& {Ledwosinska}}]{difran08}
{Di Francesco}, J., {Johnstone}, D., {Kirk}, H., {MacKenzie}, T., \&
  {Ledwosinska}, E. 2008, \apjs, 175, 277

\bibitem[{{Dobbs} {et~al.}(2005){Dobbs}, {Bonnell}, \& {Clark}}]{Dobbs05}
{Dobbs}, C.~L., {Bonnell}, I.~A., \& {Clark}, P.~C. 2005, \mnras, 360, 2

\bibitem[{{Elmegreen}(2002)}]{Elmegreen02}
{Elmegreen}, B.~G. 2002, \apj, 577, 206

\bibitem[{{Enoch} {et~al.}(2009){Enoch}, {Evans}, {Sargent}, \&
  {Glenn}}]{Enoch08young}
{Enoch}, M.~L., {Evans}, II, N.~J., {Sargent}, A.~I., \& {Glenn}, J. 2009,
  \apj, 692, 973

\bibitem[{{Enoch} {et~al.}(2008){Enoch}, {Evans}, {Sargent}, {Glenn},
  {Rosolowsky}, \& {Myers}}]{Enoch08lifetimes}
{Enoch}, M.~L., {Evans}, II, N.~J., {Sargent}, A.~I., {Glenn}, J.,
  {Rosolowsky}, E., \& {Myers}, P. 2008, \apj, 684, 1240

\bibitem[{{Evans} {et~al.}(2003){Evans}, {Allen}, {Blake}, {Boogert}, {Bourke},
  {Harvey}, {Kessler}, {Koerner}, {Lee}, {Mundy}, {Myers}, {Padgett},
  {Pontoppidan}, {Sargent}, {Stapelfeldt}, {van Dishoeck}, {Young}, \&
  {Young}}]{Evans03}
{Evans}, II, N.~J., {Allen}, L.~E., {Blake}, G.~A., {Boogert}, A.~C.~A.,
  {Bourke}, T., {Harvey}, P.~M., {Kessler}, J.~E., {Koerner}, D.~W., {Lee},
  C.~W., {Mundy}, L.~G., {Myers}, P.~C., {Padgett}, D.~L., {Pontoppidan}, K.,
  {Sargent}, A.~I., {Stapelfeldt}, K.~R., {van Dishoeck}, E.~F., {Young},
  C.~H., \& {Young}, K.~E. 2003, \pasp, 115, 965

\bibitem[{{Evans} {et~al.}(2009){Evans}, {Dunham}, {J{\o}rgensen}, {Enoch},
  {Mer{\'{\i}}n}, {van Dishoeck}, {Alcal{\'a}}, {Myers}, {Stapelfeldt},
  {Huard}, {Allen}, {Harvey}, {van Kempen}, {Blake}, {Koerner}, {Mundy},
  {Padgett}, \& {Sargent}}]{Evans09}
{Evans}, II, N.~J., {Dunham}, M.~M., {J{\o}rgensen}, J.~K., {Enoch}, M.~L.,
  {Mer{\'{\i}}n}, B., {van Dishoeck}, E.~F., {Alcal{\'a}}, J.~M., {Myers},
  P.~C., {Stapelfeldt}, K.~R., {Huard}, T.~L., {Allen}, L.~E., {Harvey}, P.~M.,
  {van Kempen}, T., {Blake}, G.~A., {Koerner}, D.~W., {Mundy}, L.~G.,
  {Padgett}, D.~L., \& {Sargent}, A.~I. 2009, \apjs, 181, 321

\bibitem[{{Fazio} {et~al.}(2004){Fazio}, {Hora}, {Allen}, {Ashby}, {Barmby},
  {Deutsch}, {Huang}, {Kleiner}, {Marengo}, {Megeath}, {Melnick}, {Pahre},
  {Patten}, {Polizotti}, {Smith}, {Taylor}, {Wang}, {Willner}, {Hoffmann},
  {Pipher}, {Forrest}, {McMurty}, {McCreight}, {McKelvey}, {McMurray}, {Koch},
  {Moseley}, {Arendt}, {Mentzell}, {Marx}, {Losch}, {Mayman}, {Eichhorn},
  {Krebs}, {Jhabvala}, {Gezari}, {Fixsen}, {Flores}, {Shakoorzadeh}, {Jungo},
  {Hakun}, {Workman}, {Karpati}, {Kichak}, {Whitley}, {Mann}, {Tollestrup},
  {Eisenhardt}, {Stern}, {Gorjian}, {Bhattacharya}, {Carey}, {Nelson},
  {Glaccum}, {Lacy}, {Lowrance}, {Laine}, {Reach}, {Stauffer}, {Surace},
  {Wilson}, {Wright}, {Hoffman}, {Domingo}, \& {Cohen}}]{IRAC}
{Fazio}, G.~G., {Hora}, J.~L., {Allen}, L.~E., {Ashby}, M.~L.~N., {Barmby}, P.,
  {Deutsch}, L.~K., {Huang}, J.-S., {Kleiner}, S., {Marengo}, M., {Megeath},
  S.~T., {Melnick}, G.~J., {Pahre}, M.~A., {Patten}, B.~M., {Polizotti}, J.,
  {Smith}, H.~A., {Taylor}, R.~S., {Wang}, Z., {Willner}, S.~P., {Hoffmann},
  W.~F., {Pipher}, J.~L., {Forrest}, W.~J., {McMurty}, C.~W., {McCreight},
  C.~R., {McKelvey}, M.~E., {McMurray}, R.~E., {Koch}, D.~G., {Moseley}, S.~H.,
  {Arendt}, R.~G., {Mentzell}, J.~E., {Marx}, C.~T., {Losch}, P., {Mayman}, P.,
  {Eichhorn}, W., {Krebs}, D., {Jhabvala}, M., {Gezari}, D.~Y., {Fixsen},
  D.~J., {Flores}, J., {Shakoorzadeh}, K., {Jungo}, R., {Hakun}, C., {Workman},
  L., {Karpati}, G., {Kichak}, R., {Whitley}, R., {Mann}, S., {Tollestrup},
  E.~V., {Eisenhardt}, P., {Stern}, D., {Gorjian}, V., {Bhattacharya}, B.,
  {Carey}, S., {Nelson}, B.~O., {Glaccum}, W.~J., {Lacy}, M., {Lowrance},
  P.~J., {Laine}, S., {Reach}, W.~T., {Stauffer}, J.~A., {Surace}, J.~A.,
  {Wilson}, G., {Wright}, E.~L., {Hoffman}, A., {Domingo}, G., \& {Cohen}, M.
  2004, \apjs, 154, 10

\bibitem[{{Friesen} {et~al.}(2009){Friesen}, {Di Francesco}, {Shirley}, \&
  {Myers}}]{Friesen09}
{Friesen}, R.~K., {Di Francesco}, J., {Shirley}, Y.~L., \& {Myers}, P.~C. 2009,
  \apj, 697, 1457

\bibitem[{{Froebrich}(2005)}]{Froebrich05}
{Froebrich}, D. 2005, \apjs, 156, 169

\bibitem[{{Geers} {et~al.}(2009){Geers}, {van Dishoeck}, {Pontoppidan},
  {Lahuis}, {Crapsi}, {Dullemond}, \& {Blake}}]{Geers09}
{Geers}, V.~C., {van Dishoeck}, E.~F., {Pontoppidan}, K.~M., {Lahuis}, F.,
  {Crapsi}, A., {Dullemond}, C.~P., \& {Blake}, G.~A. 2009, \aap, 495, 837

\bibitem[{{Goldsmith}(2001)}]{Goldsmith01}
{Goldsmith}, P.~F. 2001, \apj, 557, 736

\bibitem[{{Goldsmith} {et~al.}(2008){Goldsmith}, {Heyer}, {Narayanan}, {Snell},
  {Li}, \& {Brunt}}]{Goldsmith08}
{Goldsmith}, P.~F., {Heyer}, M., {Narayanan}, G., {Snell}, R., {Li}, D., \&
  {Brunt}, C. 2008, \apj, 680, 428

\bibitem[{{Gutermuth} {et~al.}(2008){Gutermuth}, {Myers}, {Megeath}, {Allen},
  {Pipher}, {Muzerolle}, {Porras}, {Winston}, \& {Fazio}}]{Gutermuth08}
{Gutermuth}, R.~A., {Myers}, P.~C., {Megeath}, S.~T., {Allen}, L.~E., {Pipher},
  J.~L., {Muzerolle}, J., {Porras}, A., {Winston}, E., \& {Fazio}, G. 2008,
  \apj, 674, 336

\bibitem[{{Hartmann}(2000)}]{Hartmann00}
{Hartmann}, L. 2000, in ESA Special Publication, Vol. 445, Star Formation from
  the Small to the Large Scale, ed. F.~{Favata}, A.~{Kaas}, \& A.~{Wilson},
  107--116

\bibitem[{{Harvey} {et~al.}(2007{\natexlab{a}}){Harvey}, {Mer{\'{\i}}n},
  {Huard}, {Rebull}, {Chapman}, {Evans}, \& {Myers}}]{Harvey07long}
{Harvey}, P., {Mer{\'{\i}}n}, B., {Huard}, T.~L., {Rebull}, L.~M., {Chapman},
  N., {Evans}, II, N.~J., \& {Myers}, P.~C. 2007{\natexlab{a}}, \apj, 663, 1149

\bibitem[{{Harvey} {et~al.}(2006){Harvey}, {Chapman}, {Lai}, {Evans}, {Allen},
  {J{\o}rgensen}, {Mundy}, {Huard}, {Porras}, {Cieza}, {Myers}, {Mer{\'{\i}}n},
  {van Dishoeck}, {Young}, {Spiesman}, {Blake}, {Koerner}, {Padgett},
  {Sargent}, \& {Stapelfeldt}}]{Harvey06}
{Harvey}, P.~M., {Chapman}, N., {Lai}, S.-P., {Evans}, II, N.~J., {Allen},
  L.~E., {J{\o}rgensen}, J.~K., {Mundy}, L.~G., {Huard}, T.~L., {Porras}, A.,
  {Cieza}, L., {Myers}, P.~C., {Mer{\'{\i}}n}, B., {van Dishoeck}, E.~F.,
  {Young}, K.~E., {Spiesman}, W., {Blake}, G.~A., {Koerner}, D.~W., {Padgett},
  D.~L., {Sargent}, A.~I., \& {Stapelfeldt}, K.~R. 2006, \apj, 644, 307

\bibitem[{{Harvey} {et~al.}(2007{\natexlab{b}}){Harvey}, {Rebull}, {Brooke},
  {Spiesman}, {Chapman}, {Huard}, {Evans}, {Cieza}, {Lai}, {Allen}, {Mundy},
  {Padgett}, {Sargent}, {Stapelfeldt}, {Myers}, {van Dishoeck}, {Blake}, \&
  {Koerner}}]{Harvey07short}
{Harvey}, P.~M., {Rebull}, L.~M., {Brooke}, T., {Spiesman}, W.~J., {Chapman},
  N., {Huard}, T.~L., {Evans}, II, N.~J., {Cieza}, L., {Lai}, S.-P., {Allen},
  L.~E., {Mundy}, L.~G., {Padgett}, D.~L., {Sargent}, A.~I., {Stapelfeldt},
  K.~R., {Myers}, P.~C., {van Dishoeck}, E.~F., {Blake}, G.~A., \& {Koerner},
  D.~W. 2007{\natexlab{b}}, \apj, 663, 1139

\bibitem[{{Hatchell} {et~al.}(2007){Hatchell}, {Fuller}, {Richer}, {Harries},
  \& {Ladd}}]{Hatchell07}
{Hatchell}, J., {Fuller}, G.~A., {Richer}, J.~S., {Harries}, T.~J., \& {Ladd},
  E.~F. 2007, \aap, 468, 1009

\bibitem[{{Hatchell} {et~al.}(2005){Hatchell}, {Richer}, {Fuller},
  {Qualtrough}, {Ladd}, \& {Chandler}}]{Hatchell05}
{Hatchell}, J., {Richer}, J.~S., {Fuller}, G.~A., {Qualtrough}, C.~J., {Ladd},
  E.~F., \& {Chandler}, C.~J. 2005, \aap, 440, 151

\bibitem[{{Henning} {et~al.}(1995){Henning}, {Michel}, \&
  {Stognienko}}]{Henning95}
{Henning}, T., {Michel}, B., \& {Stognienko}, R. 1995, \planss, 43, 1333

\bibitem[{{Isobe} {et~al.}(1990){Isobe}, {Feigelson}, {Akritas}, \&
  {Babu}}]{IsobeSixlin}
{Isobe}, T., {Feigelson}, E.~D., {Akritas}, M.~G., \& {Babu}, G.~J. 1990, \apj,
  364, 104

\bibitem[{{Johnstone} {et~al.}(2004){Johnstone}, {Di Francesco}, \&
  {Kirk}}]{Johnstone04}
{Johnstone}, D., {Di Francesco}, J., \& {Kirk}, H. 2004, \apjl, 611, L45

\bibitem[{{Johnstone} {et~al.}(2000{\natexlab{a}}){Johnstone}, {Wilson},
  {Moriarty-Schieven}, {Giannakopoulou-Creighton}, \&
  {Gregersen}}]{Johnstone00matrix}
{Johnstone}, D., {Wilson}, C.~D., {Moriarty-Schieven}, G.,
  {Giannakopoulou-Creighton}, J., \& {Gregersen}, E. 2000{\natexlab{a}}, \apjs,
  131, 505

\bibitem[{{Johnstone} {et~al.}(2000{\natexlab{b}}){Johnstone}, {Wilson},
  {Moriarty-Schieven}, {Joncas}, {Smith}, {Gregersen}, \& {Fich}}]{Johnstone00}
{Johnstone}, D., {Wilson}, C.~D., {Moriarty-Schieven}, G., {Joncas}, G.,
  {Smith}, G., {Gregersen}, E., \& {Fich}, M. 2000{\natexlab{b}}, \apj, 545,
  327

\bibitem[{{J{\o}rgensen} {et~al.}(2006){J{\o}rgensen}, {Harvey}, {Evans},
  {Huard}, {Allen}, {Porras}, {Blake}, {Bourke}, {Chapman}, {Cieza}, {Koerner},
  {Lai}, {Mundy}, {Myers}, {Padgett}, {Rebull}, {Sargent}, {Spiesman},
  {Stapelfeldt}, {van Dishoeck}, {Wahhaj}, \& {Young}}]{Jorgensen06}
{J{\o}rgensen}, J.~K., {Harvey}, P.~M., {Evans}, II, N.~J., {Huard}, T.~L.,
  {Allen}, L.~E., {Porras}, A., {Blake}, G.~A., {Bourke}, T.~L., {Chapman}, N.,
  {Cieza}, L., {Koerner}, D.~W., {Lai}, S.-P., {Mundy}, L.~G., {Myers}, P.~C.,
  {Padgett}, D.~L., {Rebull}, L., {Sargent}, A.~I., {Spiesman}, W.,
  {Stapelfeldt}, K.~R., {van Dishoeck}, E.~F., {Wahhaj}, Z., \& {Young}, K.~E.
  2006, \apj, 645, 1246

\bibitem[{{J{\o}rgensen} {et~al.}(2007){J{\o}rgensen}, {Johnstone}, {Kirk}, \&
  {Myers}}]{Jorgensen07}
{J{\o}rgensen}, J.~K., {Johnstone}, D., {Kirk}, H., \& {Myers}, P.~C. 2007,
  \apj, 656, 293

\bibitem[{{J{\o}rgensen} {et~al.}(2008){J{\o}rgensen}, {Johnstone}, {Kirk},
  {Myers}, {Allen}, \& {Shirley}}]{Jorgensen08}
{J{\o}rgensen}, J.~K., {Johnstone}, D., {Kirk}, H., {Myers}, P.~C., {Allen},
  L.~E., \& {Shirley}, Y.~L. 2008, \apj, 683, 822

\bibitem[{{Kirk} {et~al.}(2006){Kirk}, {Johnstone}, \& {Di Francesco}}]{Kirk06}
{Kirk}, H., {Johnstone}, D., \& {Di Francesco}, J. 2006, \apj, 646, 1009

\bibitem[{{Kirk} {et~al.}(2009){Kirk}, {Ward-Thompson}, {Di Francesco},
  {Bourke}, {Evans}, {Mer{\'{\i}}n}, {Allen}, {Cieza}, {Dunham}, {Harvey},
  {Huard}, {J{\o}rgensen}, {Miller}, {Noriega-Crespo}, {Peterson}, {Ray}, \&
  {Rebull}}]{JKirk09}
{Kirk}, J.~M., {Ward-Thompson}, D., {Di Francesco}, J., {Bourke}, T.~L.,
  {Evans}, N.~J., {Mer{\'{\i}}n}, B., {Allen}, L.~E., {Cieza}, L.~A., {Dunham},
  M.~M., {Harvey}, P., {Huard}, T., {J{\o}rgensen}, J.~K., {Miller}, J.~F.,
  {Noriega-Crespo}, A., {Peterson}, D., {Ray}, T.~P., \& {Rebull}, L.~M. 2009,
  \apjs, 185, 198

\bibitem[{{Kurtz} {et~al.}(2000){Kurtz}, {Cesaroni}, {Churchwell}, {Hofner}, \&
  {Walmsley}}]{Kurtz00}
{Kurtz}, S., {Cesaroni}, R., {Churchwell}, E., {Hofner}, P., \& {Walmsley},
  C.~M. 2000, Protostars and Planets IV, 299

\bibitem[{{Lada} {et~al.}(1994){Lada}, {Lada}, {Clemens}, \& {Bally}}]{Lada94}
{Lada}, C.~J., {Lada}, E.~A., {Clemens}, D.~P., \& {Bally}, J. 1994, \apj, 429,
  694

\bibitem[{{Lombardi} \& {Alves}(2001)}]{Lombardi01}
{Lombardi}, M., \& {Alves}, J. 2001, \aap, 377, 1023

\bibitem[{{Megeath} {et~al.}(2004){Megeath}, {Allen}, {Gutermuth}, {Pipher},
  {Myers}, {Calvet}, {Hartmann}, {Muzerolle}, \& {Fazio}}]{Megeath04}
{Megeath}, S.~T., {Allen}, L.~E., {Gutermuth}, R.~A., {Pipher}, J.~L., {Myers},
  P.~C., {Calvet}, N., {Hartmann}, L., {Muzerolle}, J., \& {Fazio}, G.~G. 2004,
  \apjs, 154, 367

\bibitem[{{Megeath} {et~al.}(2009){Megeath}, {Allgaier}, {Young}, {Allen},
  {Pipher}, \& {Wilson}}]{Megeath09}
{Megeath}, S.~T., {Allgaier}, E., {Young}, E., {Allen}, T., {Pipher}, J.~L., \&
  {Wilson}, T.~L. 2009, \aj, 137, 4072

\bibitem[{{Motte} {et~al.}(1998){Motte}, {Andre}, \& {Neri}}]{Motte98}
{Motte}, F., {Andre}, P., \& {Neri}, R. 1998, \aap, 336, 150

\bibitem[{{Myers}(2008)}]{Myers08}
{Myers}, P.~C. 2008, \apj, 687, 340

\bibitem[{{Nutter} {et~al.}(2008){Nutter}, {Kirk}, {Stamatellos}, \&
  {Ward-Thompson}}]{Nutter08}
{Nutter}, D., {Kirk}, J.~M., {Stamatellos}, D., \& {Ward-Thompson}, D. 2008,
  \mnras, 384, 755

\bibitem[{{Nutter} \& {Ward-Thompson}(2007)}]{Nutter07}
{Nutter}, D., \& {Ward-Thompson}, D. 2007, \mnras, 374, 1413

\bibitem[{{Offner} {et~al.}(2009){Offner}, {Klein}, {McKee}, \&
  {Krumholz}}]{Offner09}
{Offner}, S.~S.~R., {Klein}, R.~I., {McKee}, C.~F., \& {Krumholz}, M.~R. 2009,
  \apj, 703, 131

\bibitem[{{Padgett} {et~al.}(2008){Padgett}, {Rebull}, {Stapelfeldt},
  {Chapman}, {Lai}, {Mundy}, {Evans}, {Brooke}, {Cieza}, {Spiesman},
  {Noriega-Crespo}, {McCabe}, {Allen}, {Blake}, {Harvey}, {Huard},
  {J{\o}rgensen}, {Koerner}, {Myers}, {Sargent}, {Teuben}, {van Dishoeck},
  {Wahhaj}, \& {Young}}]{Padgett08}
{Padgett}, D.~L., {Rebull}, L.~M., {Stapelfeldt}, K.~R., {Chapman}, N.~L.,
  {Lai}, S.-P., {Mundy}, L.~G., {Evans}, II, N.~J., {Brooke}, T.~Y., {Cieza},
  L.~A., {Spiesman}, W.~J., {Noriega-Crespo}, A., {McCabe}, C.-E., {Allen},
  L.~E., {Blake}, G.~A., {Harvey}, P.~M., {Huard}, T.~L., {J{\o}rgensen},
  J.~K., {Koerner}, D.~W., {Myers}, P.~C., {Sargent}, A.~I., {Teuben}, P., {van
  Dishoeck}, E.~F., {Wahhaj}, Z., \& {Young}, K.~E. 2008, \apj, 672, 1013

\bibitem[{{Peterson} \& {Megeath}(2008)}]{PetersonMegeath08}
{Peterson}, D.~E., \& {Megeath}, S.~T. 2008, {The Orion Molecular Cloud 2/3 and
  NGC 1977 Regions} (Handbook of Star Forming Regions, Volume I: The Northern
  Sky ASP Monograph Publications, Vol.~4.~Edited by Bo Reipurth, p.590-620)

\bibitem[{{Porras} {et~al.}(2007){Porras}, {J{\o}rgensen}, {Allen}, {Evans},
  {Bourke}, {Alcal{\'a}}, {Dunham}, {Blake}, {Chapman}, {Cieza}, {Harvey},
  {Huard}, {Koerner}, {Mundy}, {Myers}, {Padgett}, {Sargent}, {Stapelfeldt},
  {Teuben}, {van Dishoeck}, {Wahhaj}, \& {Young}}]{Porras07}
{Porras}, A., {J{\o}rgensen}, J.~K., {Allen}, L.~E., {Evans}, II, N.~J.,
  {Bourke}, T.~L., {Alcal{\'a}}, J.~M., {Dunham}, M.~M., {Blake}, G.~A.,
  {Chapman}, N., {Cieza}, L., {Harvey}, P.~M., {Huard}, T.~L., {Koerner},
  D.~W., {Mundy}, L.~G., {Myers}, P.~C., {Padgett}, D.~L., {Sargent}, A.~I.,
  {Stapelfeldt}, K.~R., {Teuben}, P., {van Dishoeck}, E.~F., {Wahhaj}, Z., \&
  {Young}, K.~E. 2007, \apj, 656, 493

\bibitem[{{Rebull} {et~al.}(2009){Rebull}, {Padgett}, {McCabe}, {Hillenbrand},
  {Stapelfeldt}, {Noriega-Crespo}, {Carey}, {Brooke}, {Huard}, {Terebey},
  {Audard}, {Monin}, {Fukagawa}, {Guedel}, {Knapp}, {Menard}, {Allen},
  {Angione}, {Baldovin-Saavedra}, {Bouvier}, {Briggs}, {Dougados}, {Evans},
  {Flagey}, {Guieu}, {Grosso}, {Glauser}, {Harvey}, {Hines}, {Latter},
  {Skinner}, {Strom}, {Tromp}, \& {Wolf}}]{Rebull09}
{Rebull}, L.~M., {Padgett}, D.~L., {McCabe}, C., {Hillenbrand}, L.~A.,
  {Stapelfeldt}, K.~R., {Noriega-Crespo}, A., {Carey}, S.~J., {Brooke}, T.,
  {Huard}, T., {Terebey}, S., {Audard}, M., {Monin}, J., {Fukagawa}, M.,
  {Guedel}, M., {Knapp}, G.~R., {Menard}, F., {Allen}, L.~E., {Angione}, J.~R.,
  {Baldovin-Saavedra}, C., {Bouvier}, J., {Briggs}, K., {Dougados}, C.,
  {Evans}, N.~J., {Flagey}, N., {Guieu}, S., {Grosso}, N., {Glauser}, A.~M.,
  {Harvey}, P., {Hines}, D., {Latter}, W.~B., {Skinner}, S.~L., {Strom}, S.,
  {Tromp}, J., \& {Wolf}, S. 2009, ArXiv astro-ph/0911.3176

\bibitem[{{Rebull} {et~al.}(2007){Rebull}, {Stapelfeldt}, {Evans},
  {J{\o}rgensen}, {Harvey}, {Brooke}, {Bourke}, {Padgett}, {Chapman}, {Lai},
  {Spiesman}, {Noriega-Crespo}, {Mer{\'{\i}}n}, {Huard}, {Allen}, {Blake},
  {Jarrett}, {Koerner}, {Mundy}, {Myers}, {Sargent}, {van Dishoeck}, {Wahhaj},
  \& {Young}}]{Rebull07}
{Rebull}, L.~M., {Stapelfeldt}, K.~R., {Evans}, II, N.~J., {J{\o}rgensen},
  J.~K., {Harvey}, P.~M., {Brooke}, T.~Y., {Bourke}, T.~L., {Padgett}, D.~L.,
  {Chapman}, N.~L., {Lai}, S.-P., {Spiesman}, W.~J., {Noriega-Crespo}, A.,
  {Mer{\'{\i}}n}, B., {Huard}, T., {Allen}, L.~E., {Blake}, G.~A., {Jarrett},
  T., {Koerner}, D.~W., {Mundy}, L.~G., {Myers}, P.~C., {Sargent}, A.~I., {van
  Dishoeck}, E.~F., {Wahhaj}, Z., \& {Young}, K.~E. 2007, \apjs, 171, 447

\bibitem[{{Rieke} {et~al.}(2004){Rieke}, {Young}, {Engelbracht}, {Kelly},
  {Low}, {Haller}, {Beeman}, {Gordon}, {Stansberry}, {Misselt}, {Cadien},
  {Morrison}, {Rivlis}, {Latter}, {Noriega-Crespo}, {Padgett}, {Stapelfeldt},
  {Hines}, {Egami}, {Muzerolle}, {Alonso-Herrero}, {Blaylock}, {Dole}, {Hinz},
  {Le Floc'h}, {Papovich}, {P{\'e}rez-Gonz{\'a}lez}, {Smith}, {Su}, {Bennett},
  {Frayer}, {Henderson}, {Lu}, {Masci}, {Pesenson}, {Rebull}, {Rho}, {Keene},
  {Stolovy}, {Wachter}, {Wheaton}, {Werner}, \& {Richards}}]{MIPS}
{Rieke}, G.~H., {Young}, E.~T., {Engelbracht}, C.~W., {Kelly}, D.~M., {Low},
  F.~J., {Haller}, E.~E., {Beeman}, J.~W., {Gordon}, K.~D., {Stansberry},
  J.~A., {Misselt}, K.~A., {Cadien}, J., {Morrison}, J.~E., {Rivlis}, G.,
  {Latter}, W.~B., {Noriega-Crespo}, A., {Padgett}, D.~L., {Stapelfeldt},
  K.~R., {Hines}, D.~C., {Egami}, E., {Muzerolle}, J., {Alonso-Herrero}, A.,
  {Blaylock}, M., {Dole}, H., {Hinz}, J.~L., {Le Floc'h}, E., {Papovich}, C.,
  {P{\'e}rez-Gonz{\'a}lez}, P.~G., {Smith}, P.~S., {Su}, K.~Y.~L., {Bennett},
  L., {Frayer}, D.~T., {Henderson}, D., {Lu}, N., {Masci}, F., {Pesenson}, M.,
  {Rebull}, L., {Rho}, J., {Keene}, J., {Stolovy}, S., {Wachter}, S.,
  {Wheaton}, W., {Werner}, M.~W., \& {Richards}, P.~L. 2004, \apjs, 154, 25

\bibitem[{{Robin} {et~al.}(2003){Robin}, {Reyl{\'e}}, {Derri{\`e}re}, \&
  {Picaud}}]{Robin03}
{Robin}, A.~C., {Reyl{\'e}}, C., {Derri{\`e}re}, S., \& {Picaud}, S. 2003,
  \aap, 409, 523

\bibitem[{{Rosolowsky} {et~al.}(2008){Rosolowsky}, {Pineda}, {Foster},
  {Borkin}, {Kauffmann}, {Caselli}, {Myers}, \& {Goodman}}]{Rosolowsky08}
{Rosolowsky}, E.~W., {Pineda}, J.~E., {Foster}, J.~B., {Borkin}, M.~A.,
  {Kauffmann}, J., {Caselli}, P., {Myers}, P.~C., \& {Goodman}, A.~A. 2008,
  \apjs, 175, 509

\bibitem[{{Salpeter}(1955)}]{Salpeter55}
{Salpeter}, E.~E. 1955, \apj, 121, 161

\bibitem[{{Schnee} {et~al.}(2005){Schnee}, {Ridge}, {Goodman}, \&
  {Li}}]{Schnee05}
{Schnee}, S.~L., {Ridge}, N.~A., {Goodman}, A.~A., \& {Li}, J.~G. 2005, \apj,
  634, 442

\bibitem[{{Schneider} {et~al.}(2006){Schneider}, {Bontemps}, {Simon}, {Jakob},
  {Motte}, {Miller}, {Kramer}, \& {Stutzki}}]{Schneider06}
{Schneider}, N., {Bontemps}, S., {Simon}, R., {Jakob}, H., {Motte}, F.,
  {Miller}, M., {Kramer}, C., \& {Stutzki}, J. 2006, \aap, 458, 855

\bibitem[{{Shu} {et~al.}(2004){Shu}, {Li}, \& {Allen}}]{Shu04}
{Shu}, F.~H., {Li}, Z.-Y., \& {Allen}, A. 2004, \apj, 601, 930

\bibitem[{{van Dishoeck} \& {van der Tak}(2000)}]{vanDishoeck00}
{van Dishoeck}, E.~F., \& {van der Tak}, F.~F.~S. 2000, in IAU Symposium, Vol.
  197, From Molecular Clouds to Planetary, ed. {Y.~C.~Minh \& E.~F.~van
  Dishoeck}, 97--112

\bibitem[{{Ward-Thompson} {et~al.}(2007{\natexlab{a}}){Ward-Thompson},
  {Andr{\'e}}, {Crutcher}, {Johnstone}, {Onishi}, \& {Wilson}}]{Ward-T07}
{Ward-Thompson}, D., {Andr{\'e}}, P., {Crutcher}, R., {Johnstone}, D.,
  {Onishi}, T., \& {Wilson}, C. 2007{\natexlab{a}}, in Protostars and Planets
  V, ed. B.~{Reipurth}, D.~{Jewitt}, \& K.~{Keil}, 33--46

\bibitem[{{Ward-Thompson} {et~al.}(2007{\natexlab{b}}){Ward-Thompson}, {Di
  Francesco}, {Hatchell}, {Hogerheijde}, {Nutter}, {Bastien}, {Basu},
  {Bonnell}, {Bowey}, {Brunt}, {Buckle}, {Butner}, {Cavanagh}, {Chrysostomou},
  {Curtis}, {Davis}, {Dent}, {van Dishoeck}, {Edmunds}, {Fich}, {Fiege},
  {Fissel}, {Friberg}, {Friesen}, {Frieswijk}, {Fuller}, {Gosling}, {Graves},
  {Greaves}, {Helmich}, {Hills}, {Holland}, {Houde}, {Jayawardhana},
  {Johnstone}, {Joncas}, {Kirk}, {Kirk}, {Knee}, {Matthews}, {Matthews},
  {Matzner}, {Moriarty-Schieven}, {Naylor}, {Padman}, {Plume}, {Rawlings},
  {Redman}, {Reid}, {Richer}, {Shipman}, {Simpson}, {Spaans}, {Stamatellos},
  {Tsamis}, {Viti}, {Weferling}, {White}, {Whitworth}, {Wouterloot}, {Yates},
  \& {Zhu}}]{Ward-T07GBS}
{Ward-Thompson}, D., {Di Francesco}, J., {Hatchell}, J., {Hogerheijde}, M.~R.,
  {Nutter}, D., {Bastien}, P., {Basu}, S., {Bonnell}, I., {Bowey}, J., {Brunt},
  C., {Buckle}, J., {Butner}, H., {Cavanagh}, B., {Chrysostomou}, A., {Curtis},
  E., {Davis}, C.~J., {Dent}, W.~R.~F., {van Dishoeck}, E., {Edmunds}, M.~G.,
  {Fich}, M., {Fiege}, J., {Fissel}, L., {Friberg}, P., {Friesen}, R.,
  {Frieswijk}, W., {Fuller}, G.~A., {Gosling}, A., {Graves}, S., {Greaves},
  J.~S., {Helmich}, F., {Hills}, R.~E., {Holland}, W.~S., {Houde}, M.,
  {Jayawardhana}, R., {Johnstone}, D., {Joncas}, G., {Kirk}, H., {Kirk}, J.~M.,
  {Knee}, L.~B.~G., {Matthews}, B., {Matthews}, H., {Matzner}, C.,
  {Moriarty-Schieven}, G.~H., {Naylor}, D., {Padman}, R., {Plume}, R.,
  {Rawlings}, J.~M.~C., {Redman}, R.~O., {Reid}, M., {Richer}, J.~S.,
  {Shipman}, R., {Simpson}, R.~J., {Spaans}, M., {Stamatellos}, D., {Tsamis},
  Y.~G., {Viti}, S., {Weferling}, B., {White}, G.~J., {Whitworth}, A.~P.,
  {Wouterloot}, J., {Yates}, J., \& {Zhu}, M. 2007{\natexlab{b}}, \pasp, 119,
  855

\bibitem[{{Warner}(1961)}]{Warner61}
{Warner}, B. 1961, \pasp, 73, 439

\bibitem[{{Werner} {et~al.}(2004){Werner}, {Roellig}, {Low}, {Rieke}, {Rieke},
  {Hoffmann}, {Young}, {Houck}, {Brandl}, {Fazio}, {Hora}, {Gehrz}, {Helou},
  {Soifer}, {Stauffer}, {Keene}, {Eisenhardt}, {Gallagher}, {Gautier}, {Irace},
  {Lawrence}, {Simmons}, {Van Cleve}, {Jura}, {Wright}, \&
  {Cruikshank}}]{Spitzer}
{Werner}, M.~W., {Roellig}, T.~L., {Low}, F.~J., {Rieke}, G.~H., {Rieke}, M.,
  {Hoffmann}, W.~F., {Young}, E., {Houck}, J.~R., {Brandl}, B., {Fazio}, G.~G.,
  {Hora}, J.~L., {Gehrz}, R.~D., {Helou}, G., {Soifer}, B.~T., {Stauffer}, J.,
  {Keene}, J., {Eisenhardt}, P., {Gallagher}, D., {Gautier}, T.~N., {Irace},
  W., {Lawrence}, C.~R., {Simmons}, L., {Van Cleve}, J.~E., {Jura}, M.,
  {Wright}, E.~L., \& {Cruikshank}, D.~P. 2004, \apjs, 154, 1

\bibitem[{{Williams} {et~al.}(1994){Williams}, {de Geus}, \&
  {Blitz}}]{Clumpfind}
{Williams}, J.~P., {de Geus}, E.~J., \& {Blitz}, L. 1994, \apj, 428, 693

\bibitem[{{Wilson} {et~al.}(1999){Wilson}, {Mauersberger}, {Gensheimer},
  {Muders}, \& {Bieging}}]{Wilson99}
{Wilson}, T.~L., {Mauersberger}, R., {Gensheimer}, P.~D., {Muders}, D., \&
  {Bieging}, J.~H. 1999, \apj, 525, 343

\bibitem[{{Winston} {et~al.}(2007){Winston}, {Megeath}, {Wolk}, {Muzerolle},
  {Gutermuth}, {Hora}, {Allen}, {Spitzbart}, {Myers}, \& {Fazio}}]{Winston07}
{Winston}, E., {Megeath}, S.~T., {Wolk}, S.~J., {Muzerolle}, J., {Gutermuth},
  R., {Hora}, J.~L., {Allen}, L.~E., {Spitzbart}, B., {Myers}, P., \& {Fazio},
  G.~G. 2007, \apj, 669, 493

\bibitem[{{Young} {et~al.}(2005){Young}, {Harvey}, {Brooke}, {Chapman},
  {Kauffmann}, {Bertoldi}, {Lai}, {Alcal{\'a}}, {Bourke}, {Spiesman}, {Allen},
  {Blake}, {Evans}, {Koerner}, {Mundy}, {Myers}, {Padgett}, {Salinas},
  {Sargent}, {Stapelfeldt}, {Teuben}, {van Dishoeck}, \& {Wahhaj}}]{Young05}
{Young}, K.~E., {Harvey}, P.~M., {Brooke}, T.~Y., {Chapman}, N., {Kauffmann},
  J., {Bertoldi}, F., {Lai}, S.-P., {Alcal{\'a}}, J., {Bourke}, T.~L.,
  {Spiesman}, W., {Allen}, L.~E., {Blake}, G.~A., {Evans}, II, N.~J.,
  {Koerner}, D.~W., {Mundy}, L.~G., {Myers}, P.~C., {Padgett}, D.~L.,
  {Salinas}, A., {Sargent}, A.~I., {Stapelfeldt}, K.~R., {Teuben}, P., {van
  Dishoeck}, E.~F., \& {Wahhaj}, Z. 2005, \apj, 628, 283

\end{thebibliography}

\pagebreak

\begin{appendix}
\section*{APPENDIX}
\section{Starless and Protostellar Core Lists}

Using the SLC and our 2MASS data, we identified cores and their locations within the bulk cloud. Tables \ref{starlessList} and \ref{protoList} give examples of the SLC and extinction information for cores in the Perseus cloud. For both tables, the core name reflects the J2000 position of the submillimeter peak and the mass is calculated using Equation \ref{massEq} with the temperatures given in Table \ref{prop}. The complete tables are available in the electronic edition.

\end{appendix}

\begin{figure}
\includegraphics[scale=0.73]{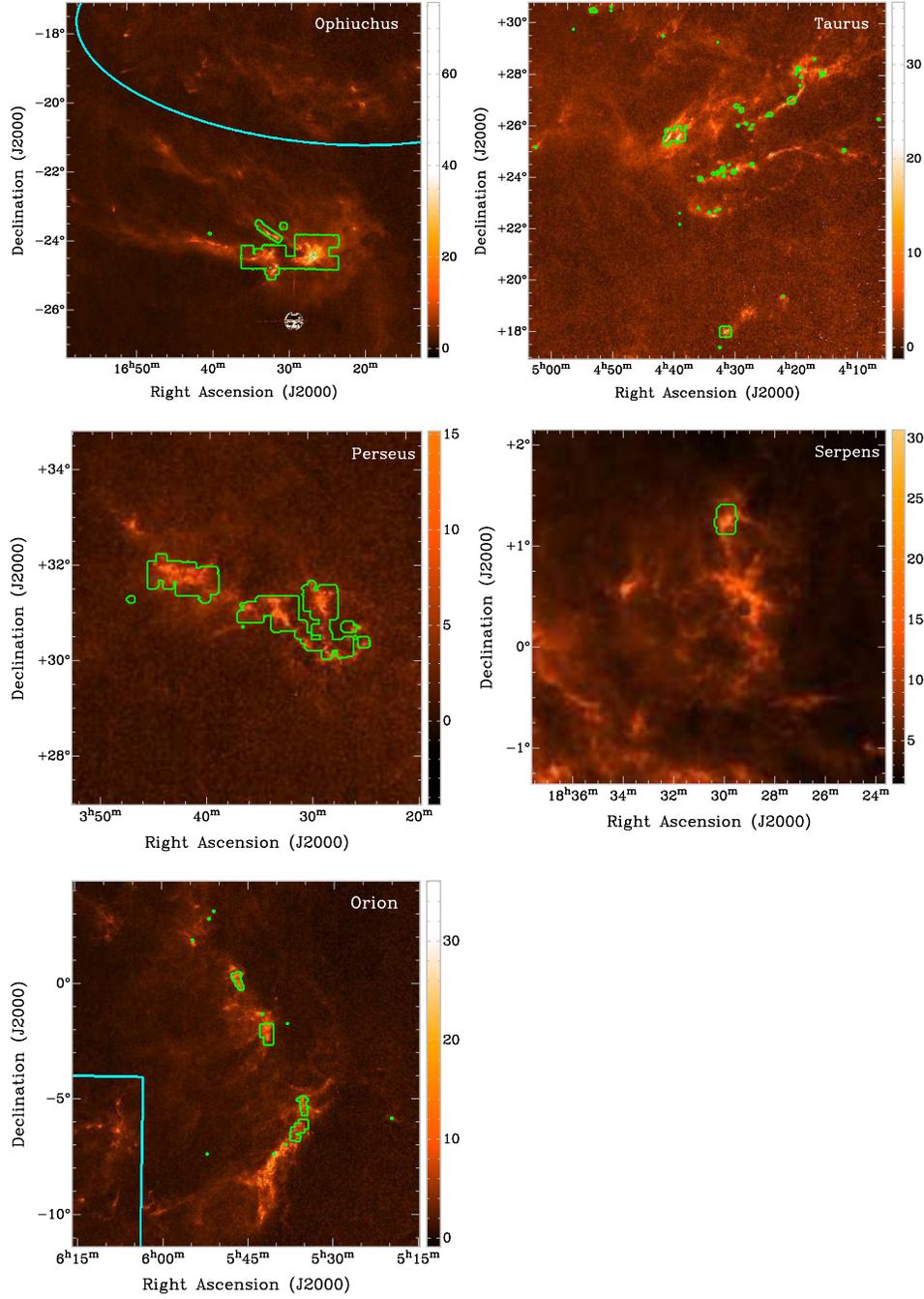}
\caption{2MASS extinction maps with Funamental SLC data (contours). For Ophiuchus and Orion, lines are also shown to mark the rough boundary between the Scorpius and Monoceros clouds, respectively. For the color figure, see the electronic edition. \label{complete}}
\end{figure}
\clearpage

\begin{figure}[h!]
\includegraphics[scale=0.9]{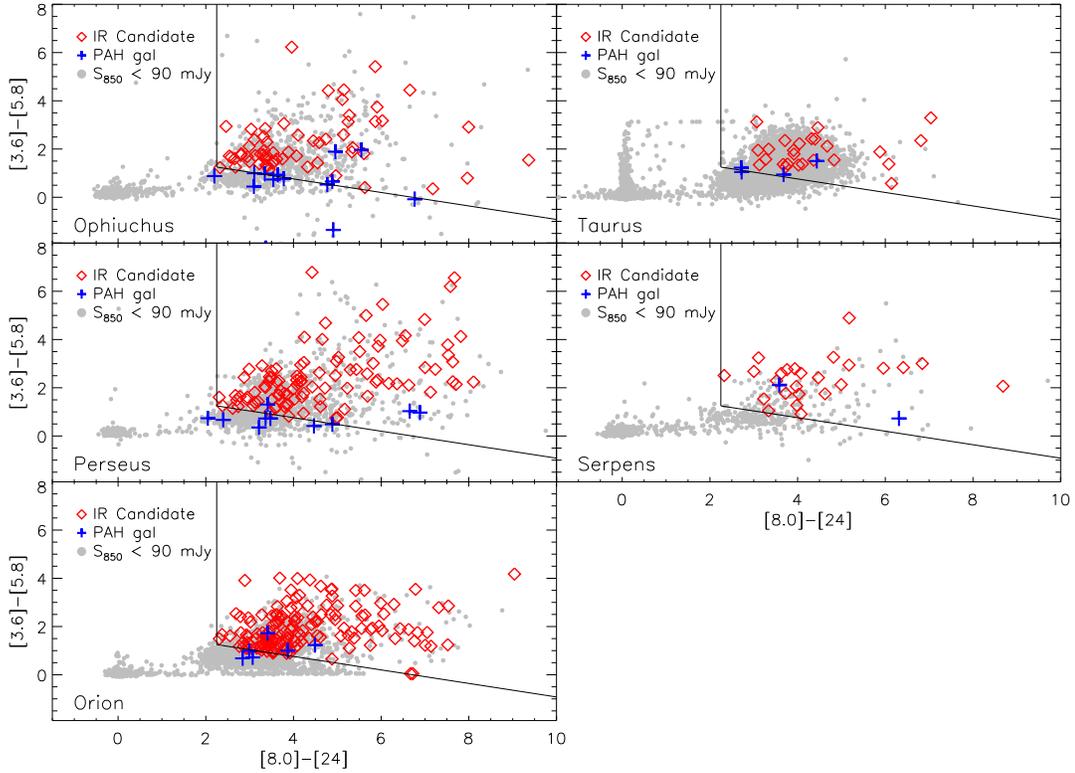}
\caption{Color-color diagrams for the Spitzer objects in our sample. The diamonds show the infrared objects that met all our color criteria. The grey circles are objects that passed the high quality conditions (CC1 and CC2) but with an associated 850 $\mu$m flux of $S_{850} < 90$ mJy beam\negExp. The crosses are objects with an associated 850 $\mu$m flux of $S_{850} > 90$ mJy beam\negExp\ but also met the colors of a star-forming PAH galaxy (CC3). The large concentration of grey points in Taurus at $[8.0]-[24] \sim 0$ are due to foreground and background stars (the c2d catalogue used a pipeline that identified and removed many stars from the catalogue). As well, a number of infrared sources in Taurus were saturated, resulting in a ``rollover'' at $[3.6]-[5.8] \sim 3$. For the color figure, see the electronic edition.\label{colorcolor}}
\end{figure}
\clearpage

\begin{figure}[h!]
\centering
\includegraphics[scale=0.6]{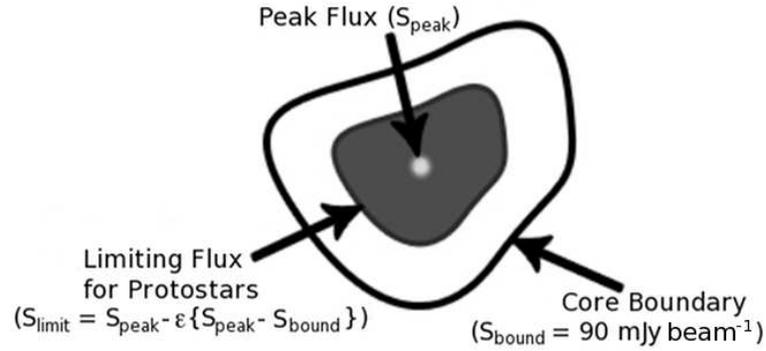}
\caption{Schematic of our classification technique for protostellar cores. A submillimeter core is considered to be protostellar if a YSO candidate identified in \S \ref{colorCrit} falls within the shaded region illustrated in this figure. The black contour represents the sensitivity boundary of the core (90 mJy beam\negExp) and the pale grey circle represents the peak intensity. The dark grey contour represents the maximum distance from the central peak for the protostar classification (as given by Equation \ref{contour}).\label{myTech}}
\end{figure}
\clearpage

\begin{figure}[h!]
\includegraphics[scale=0.92]{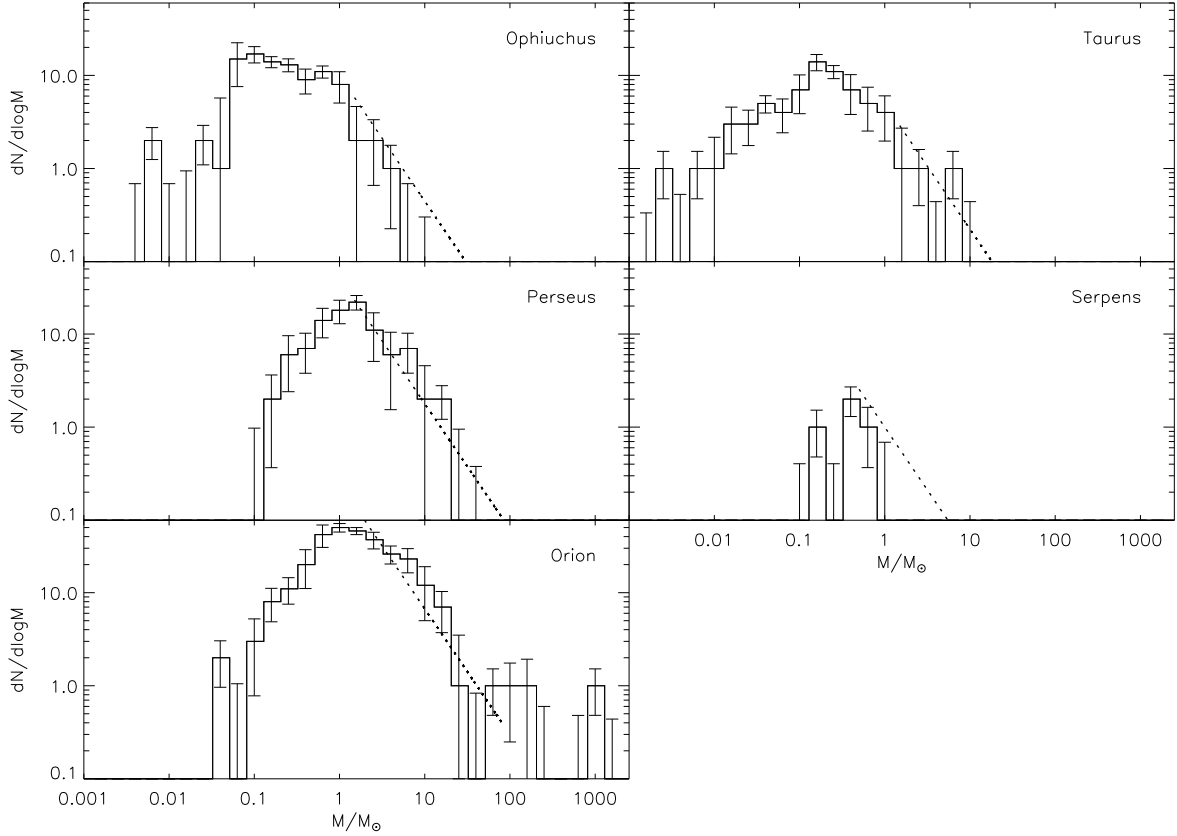}
\caption{Starless CMFs for Ophiuchus, Taurus, Perseus, Serpens, and Orion with Salpeter power-law slopes $dN/d(\log{m}) \propto m^{-1.35}$. Uncertainties were determined from varying the temperature (see text). Starless cores were identified using our classification technique outlined in \S \ref{colorCrit} and \S \ref{fluxCrit}. \label{CMF+IMF}}
\end{figure}
\clearpage

\begin{figure}[h!]
\includegraphics[scale=0.7]{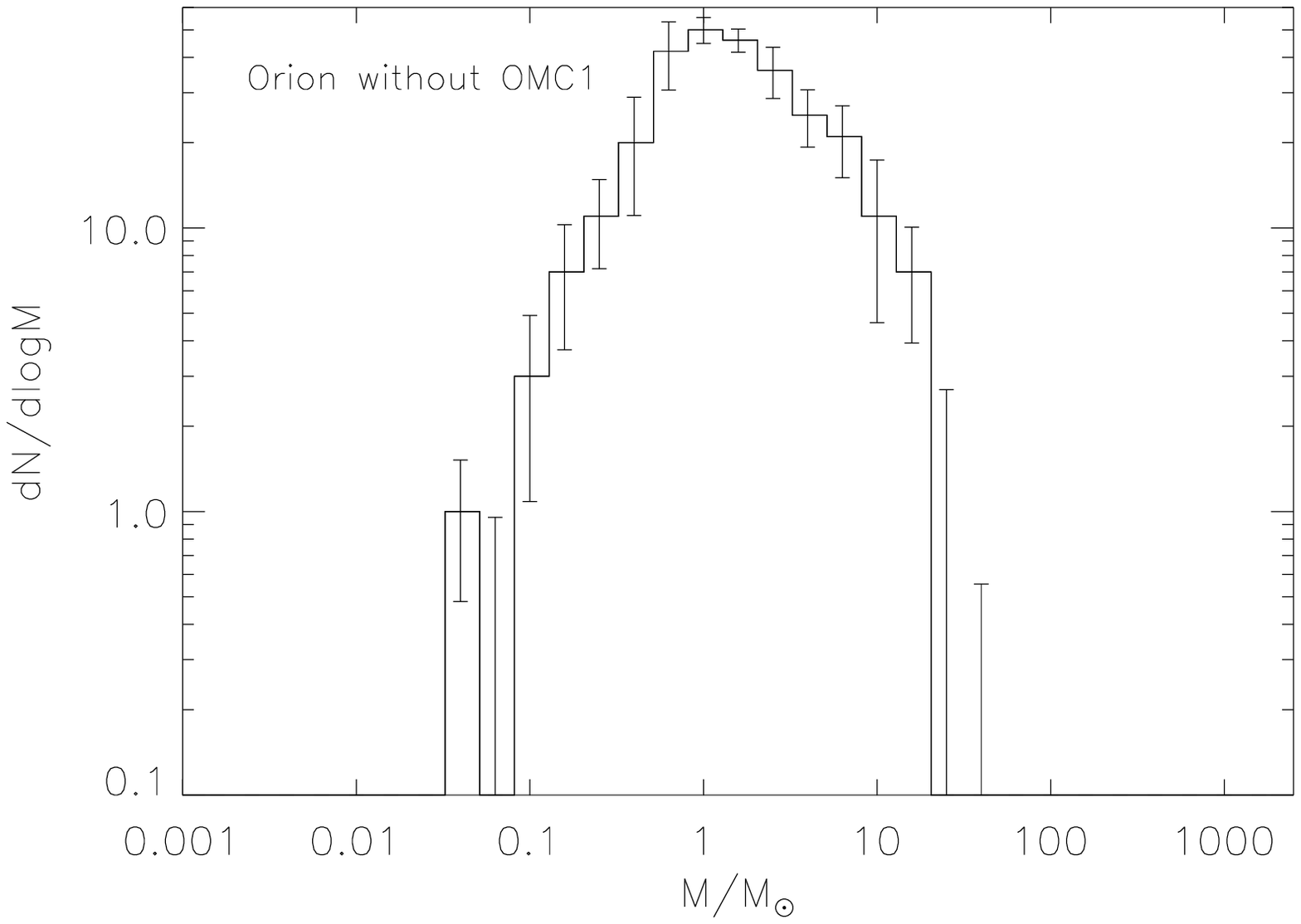}
\caption{Same as Orion in Figure \ref{CMF+IMF} but with cores towards OMC-1 removed. We used coordinates of [$5^h35^m00^s, 5^h35^m32^s$] in right ascension and [$-5\degree29\arcmin00\arcsec, -5\degree17\arcmin00\arcsec$] in declination as our boundary for OMC-1. \label{CMF-OMC1}}
\end{figure}
\clearpage

\begin{figure}[h!]
\includegraphics[scale=0.92]{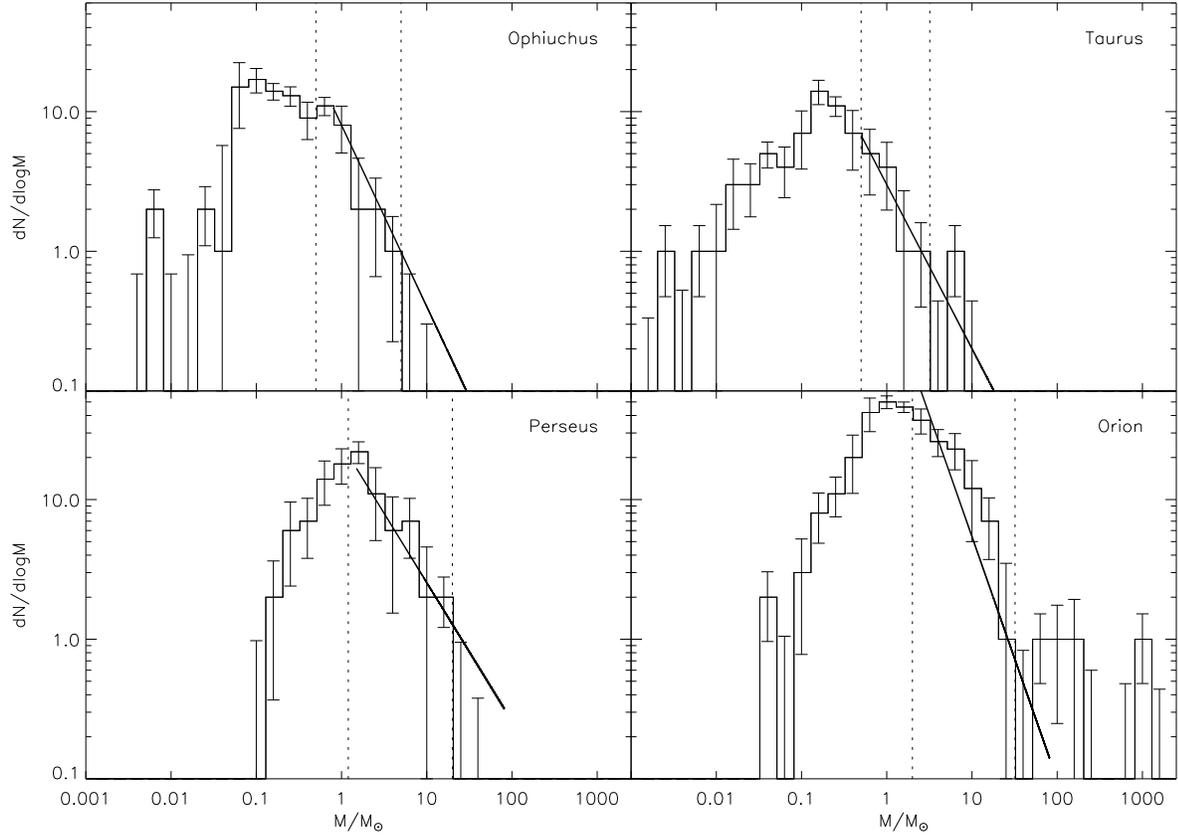}
\caption{Best fit slopes from ordinary least squares regression. The dotted lines indicate one mass range used to calculate the slope (see Table \ref{bestFitResults}). The Serpens CMF had too few mass bins to calculate the best fit slope.\label{bestFit}}
\end{figure}
\clearpage

\begin{figure}[h!]
\includegraphics[angle=-90,scale=0.75]{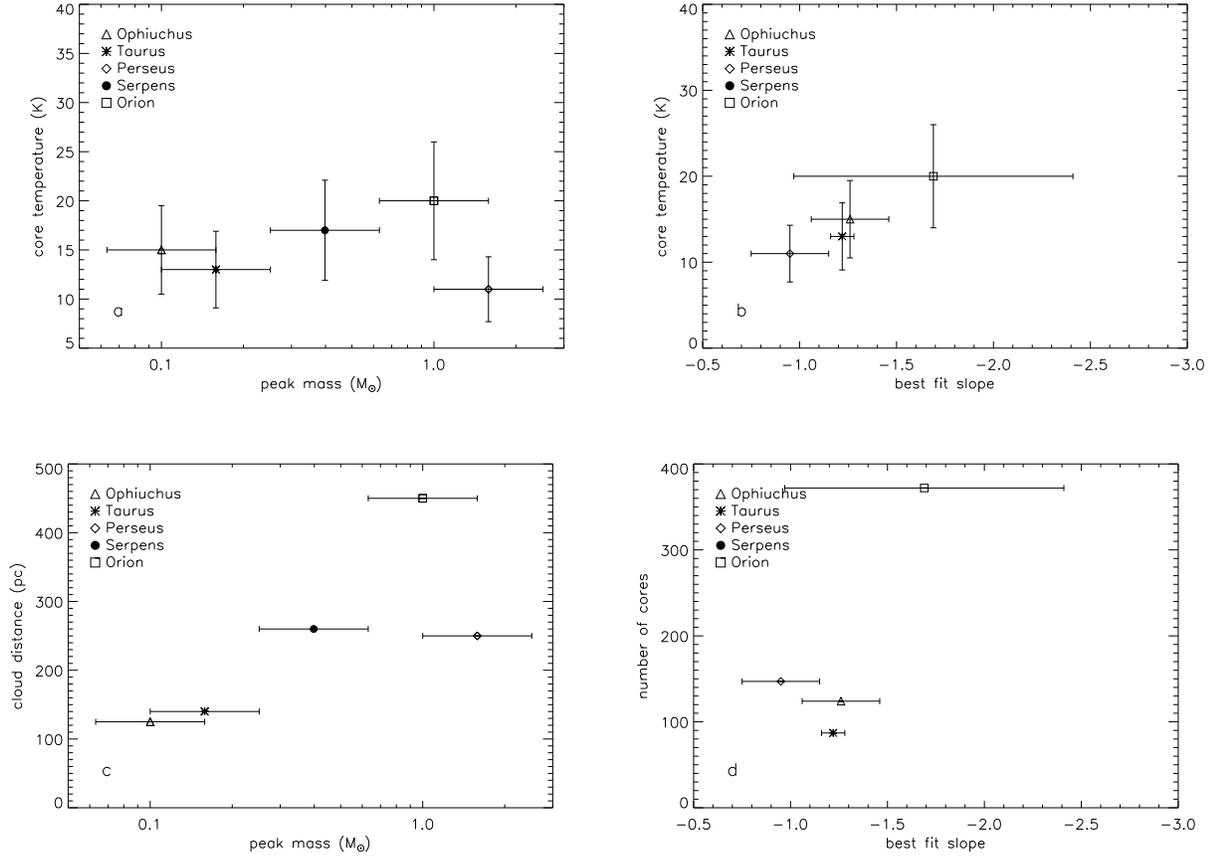}
\caption{Examples of comparisons between cloud or core properties with CMF properties. Peak masses were defined as the most populated mass bins in the starless CMFs. The best fit slopes are given in Table \ref{bestFitResults}. Serpens had too few points to determine a linear best fit slope.\label{trends}}
\end{figure}
\clearpage

\begin{figure}[h!]
\includegraphics[scale=0.88]{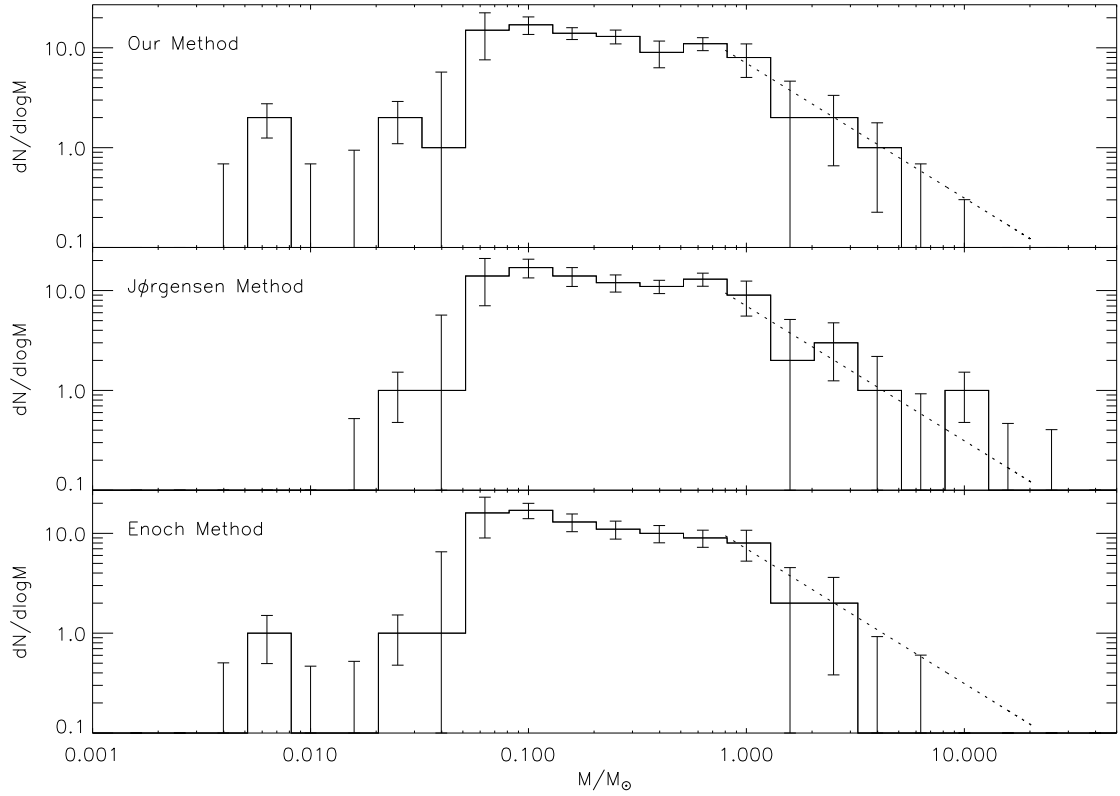}
\caption{Comparison of the resulting CMFs from applying the E08, J07 and our own core classification techniques for Ophiuchus. The dotted line represents a Salpeter power-law slope. Uncertainties were determined by varying the temperature by $\sim 30$\%. \label{CMFcompareOph}}
\end{figure}
\clearpage

\begin{figure}[h!]
\includegraphics[scale=0.88]{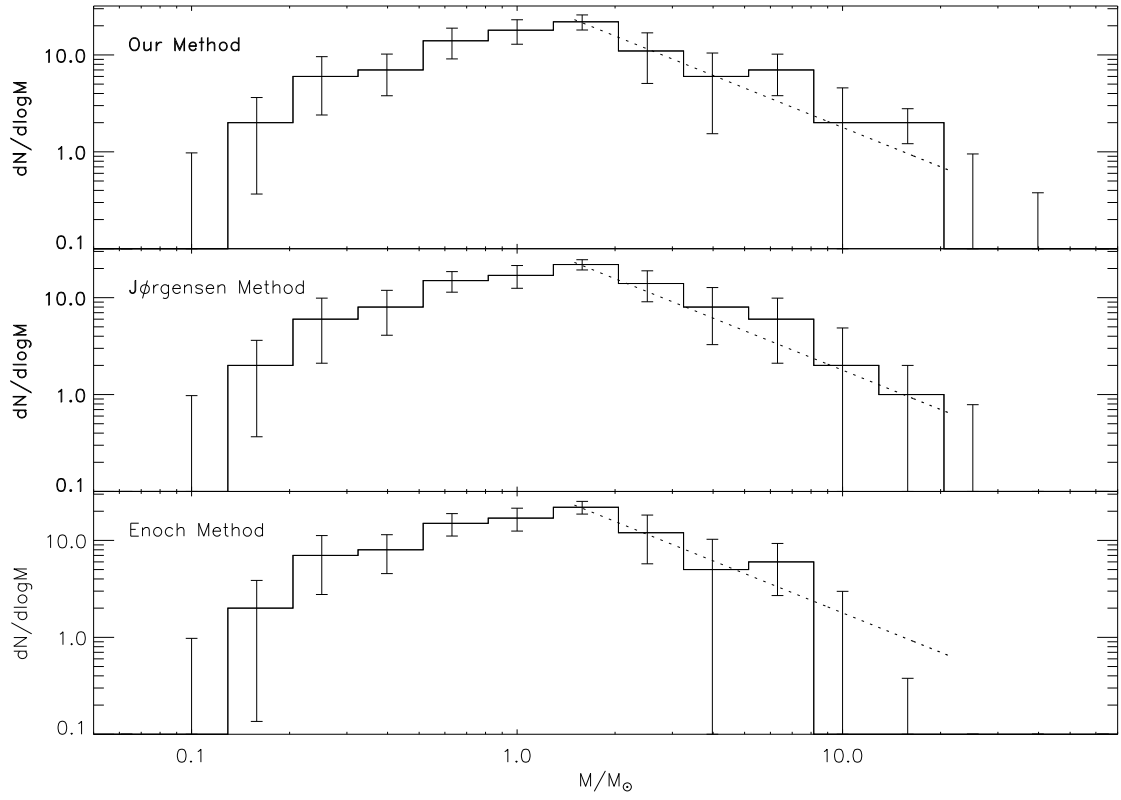}
\caption{Same as Figure \ref{CMFcompareOph} but for Perseus. \label{CMFcomparePer}}
\end{figure}
\clearpage

\begin{figure}[h!]
\includegraphics[scale=0.92]{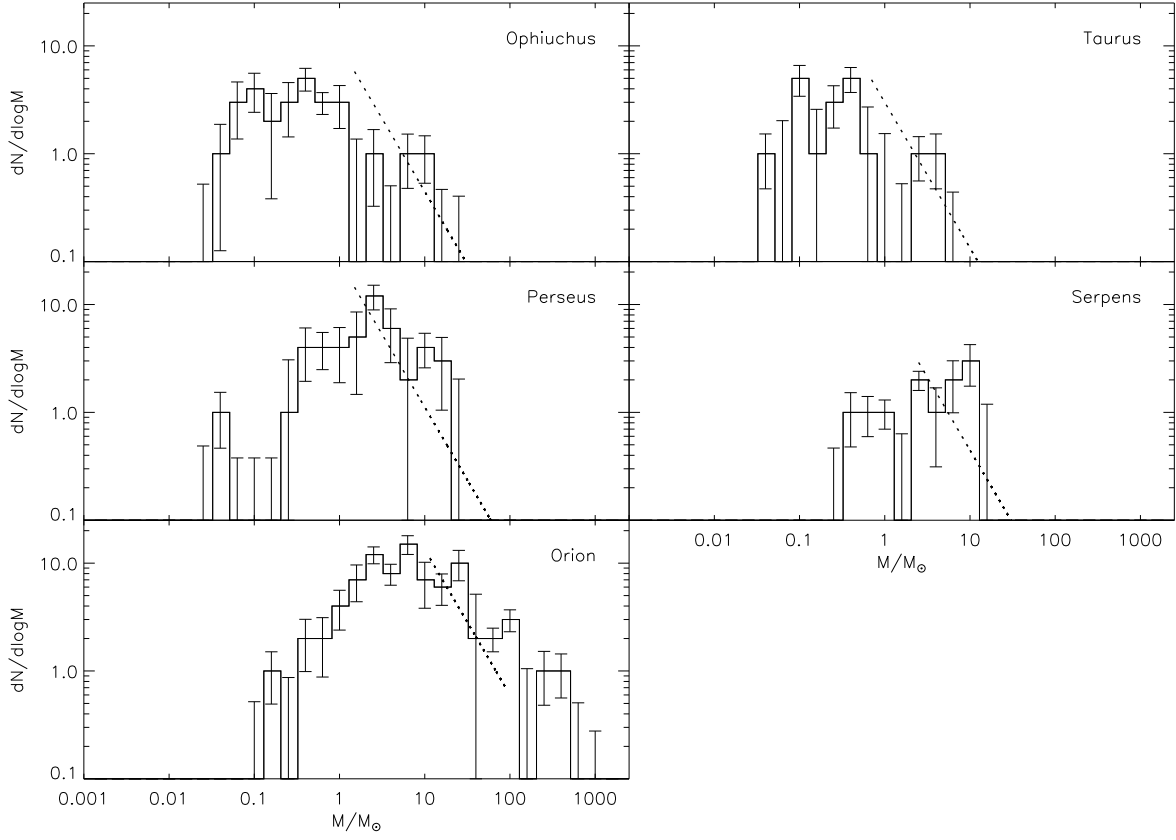}
\caption{Protostellar CMFs for all five clouds with Salpeter power-law slopes $dN/d\log{m} \propto m^{-1.35}$. Uncertainties were determined from varying the temperature up to 30\%. We assume the same temperature as the starless cores. \label{protoCMF}}
\end{figure}
\clearpage

\begin{figure}[h!]
\includegraphics[scale=0.92]{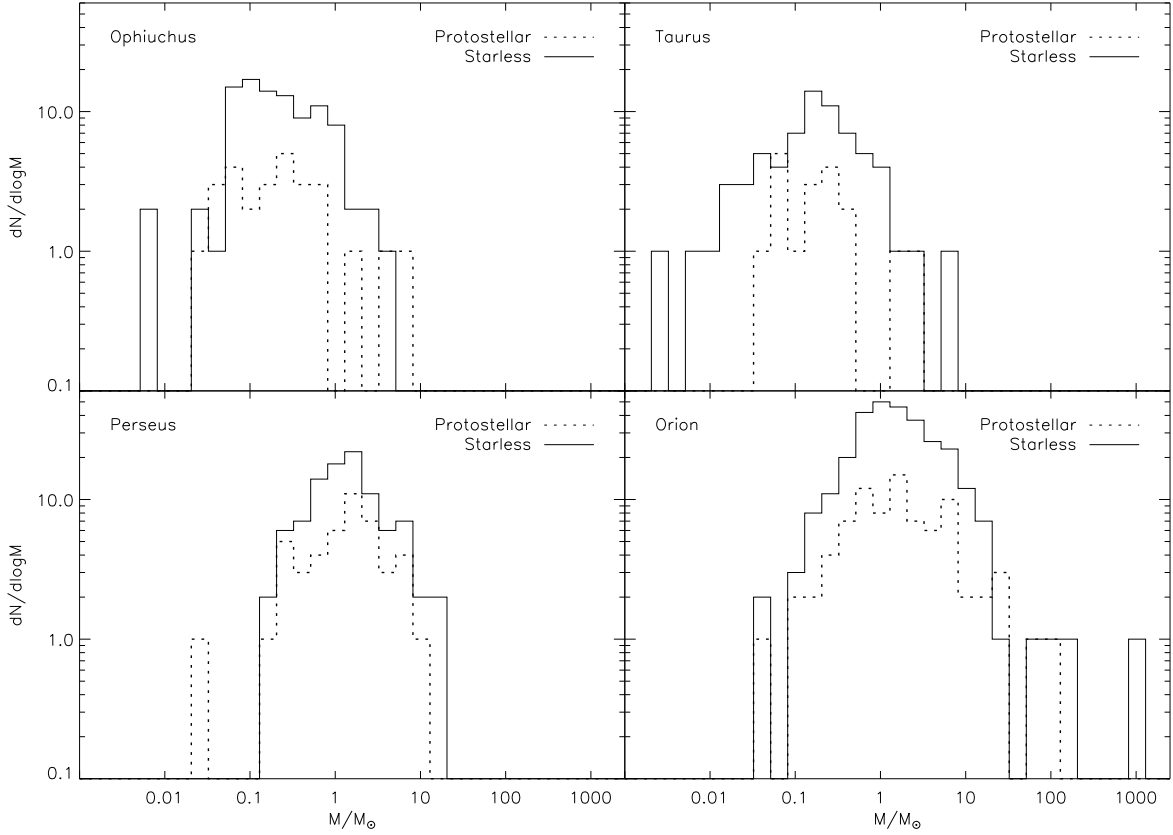}
\caption{Protostellar (dotted) and starless (solid) CMFs for all five clouds. The temperatures for the protostellar CMFs have been adjusted to match the width and peak of the starless CMF. The temperatures used are 20 K, 17 K, 15 K, and 59 K for Ophiuchus, Taurus, Perseus, and Orion protostellar CMFs, respectively. The respective starless core temperatures are 15 K, 13 K, 11 K, and 20 K. Serpens had too few cores to make this comparison. \label{CMFcompare}}
\end{figure}
\clearpage

\begin{figure}
\includegraphics[scale=0.92]{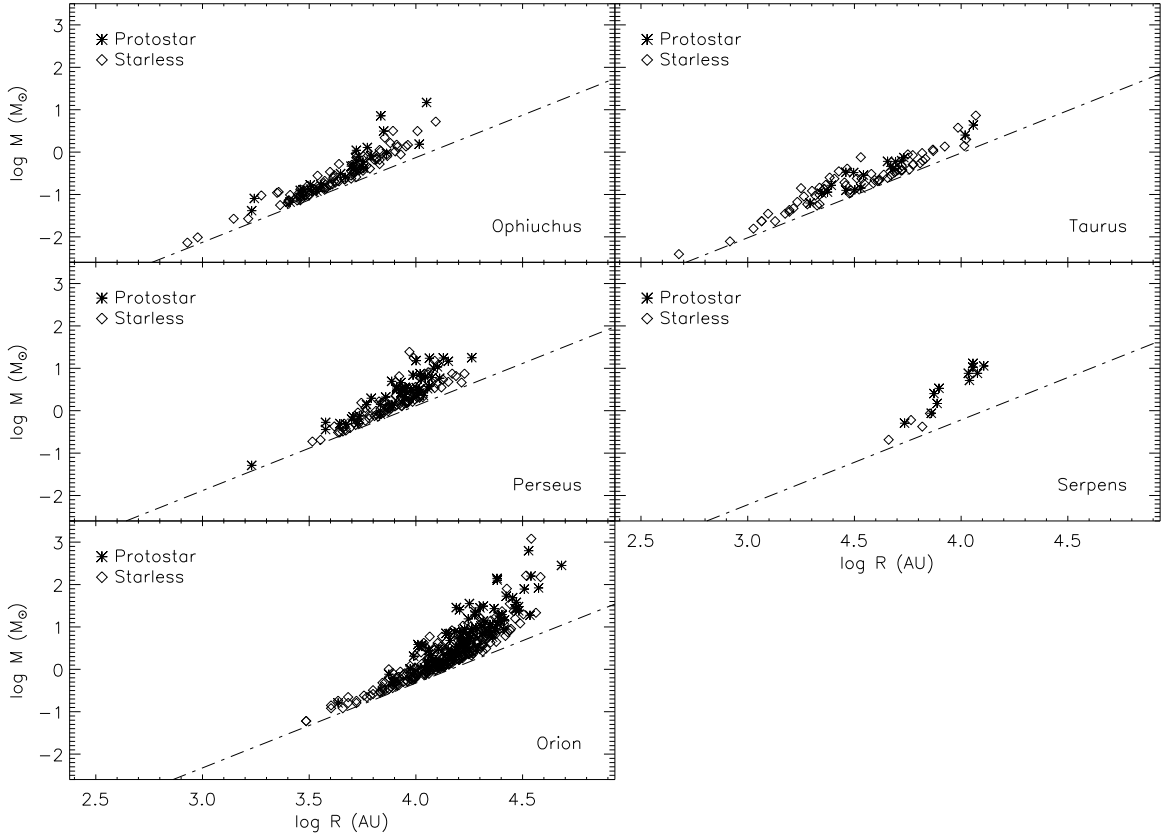}
\caption{Comparison of radius and mass for cores in the five clouds in our sample. Protostellar cores were classified using our technique outlined in \S \ref{colorCrit} and \S \ref{fluxCrit}. We assume a constant dust temperature (see Table \ref{prop}) for cores within each cloud regardless of whether the cores are protostars or starless. The threshold sensitivity of 90 mJy beam\negExp\ is shown as a dot-dashed line. \label{BEmodels}}
\end{figure}
\clearpage

\begin{figure}
\includegraphics[scale=0.92]{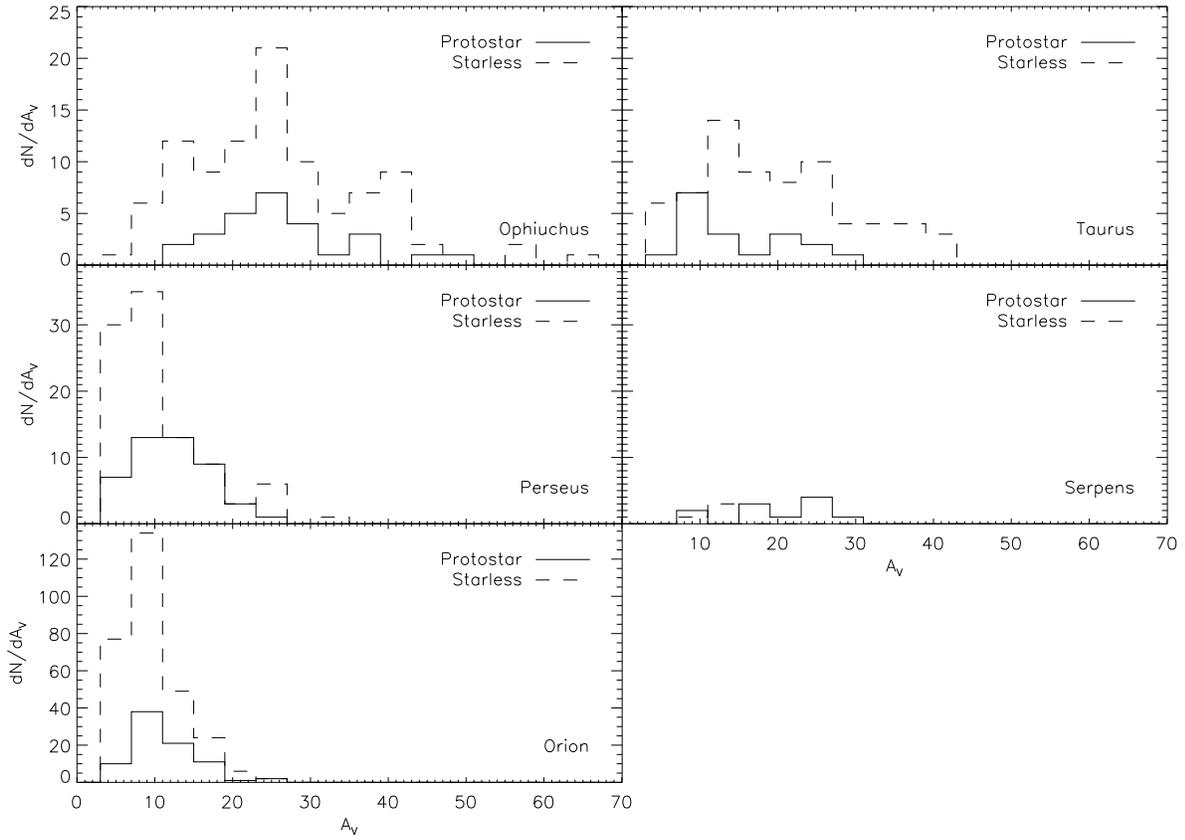}
\caption{Distribution of extinction for each core in our five clouds. Cores were classified into starless or protostellar using the technique we developed (see \S \ref{colorCrit} and \S \ref{fluxCrit}). \label{AV-dist}}
\end{figure}
\clearpage

\begin{figure}
\includegraphics[scale=0.92]{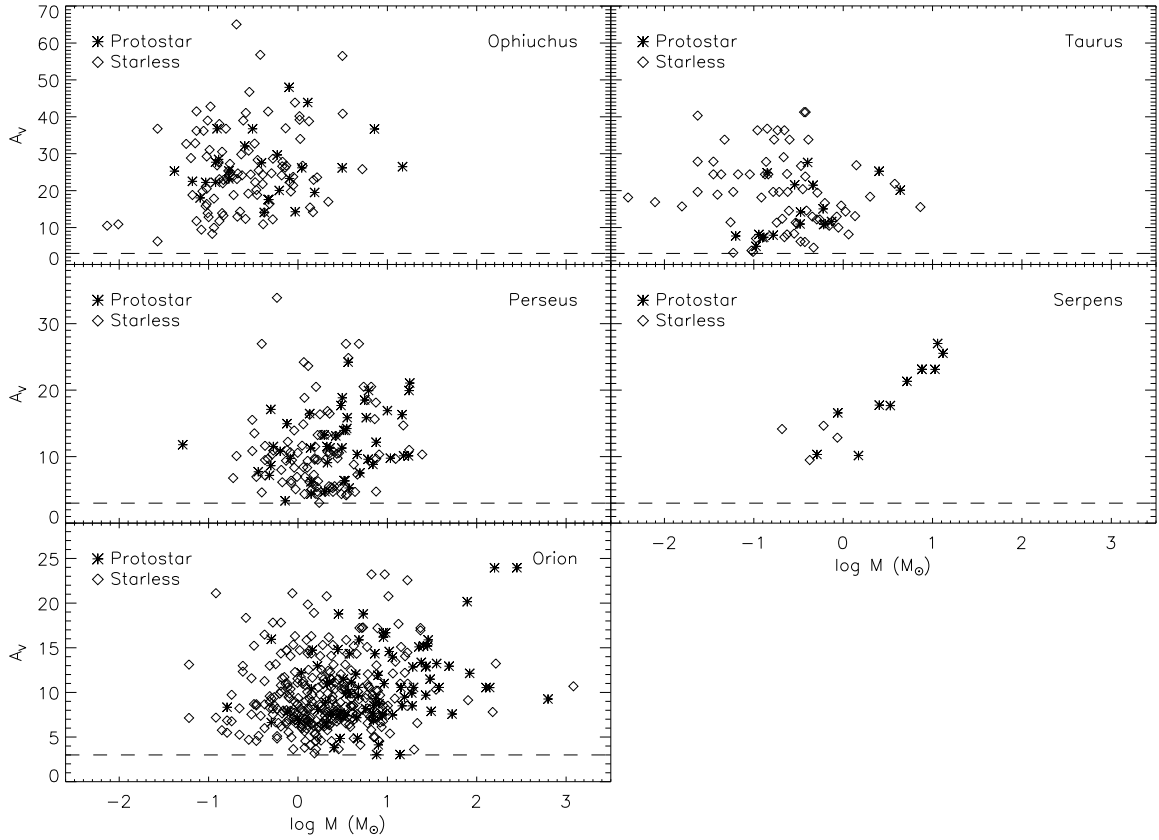}
\caption{Core extinction versus mass for starless and protostellar cores. Cores were classified using the technique outlined in \S \ref{colorCrit} and \S \ref{fluxCrit}. The dashed line illustrates the $A_V = 3$ limit we imposed on all cores.\label{AV-Mass}}
\end{figure}
\clearpage

\begin{figure}[h!]
\includegraphics[scale=0.92]{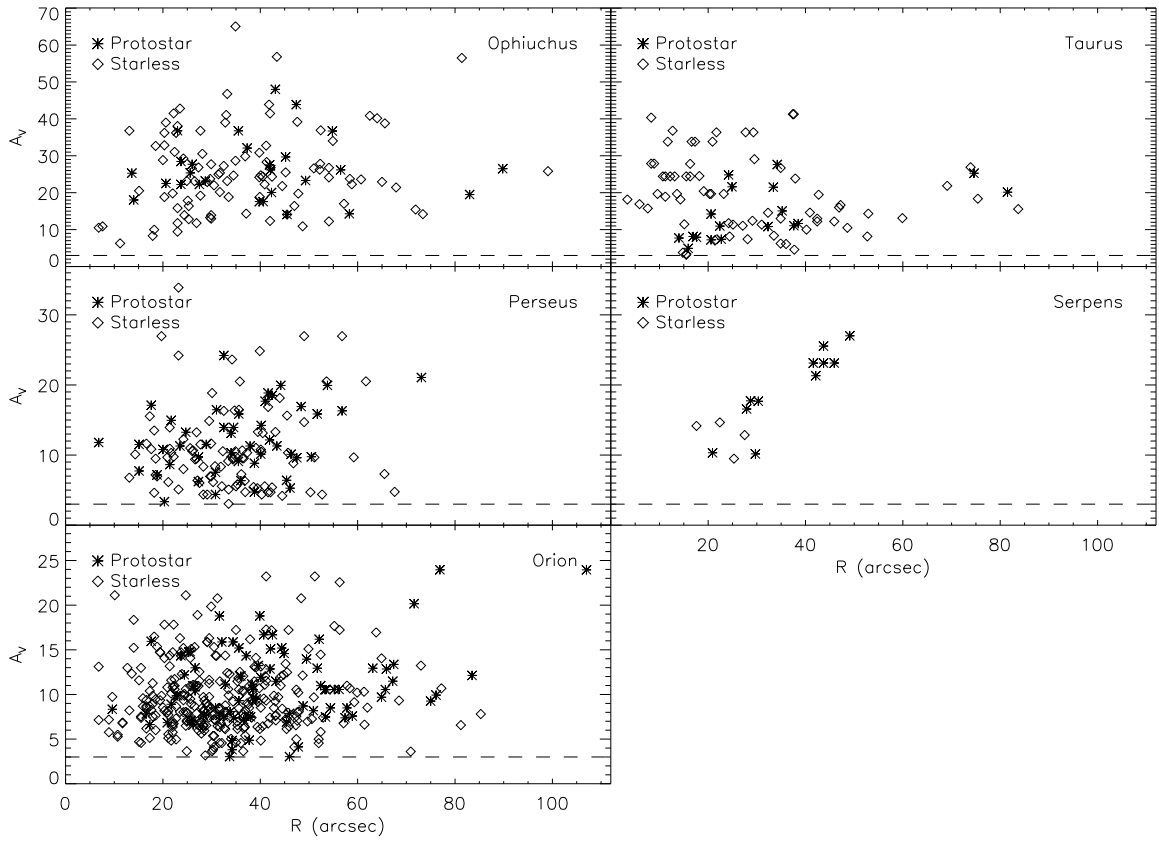}
\caption{Same as Figure \ref{AV-Mass} but showing core extinction versus size. \label{AV-Rad}}
\end{figure}
\clearpage

\begin{figure}[h!]
\includegraphics[scale=0.92]{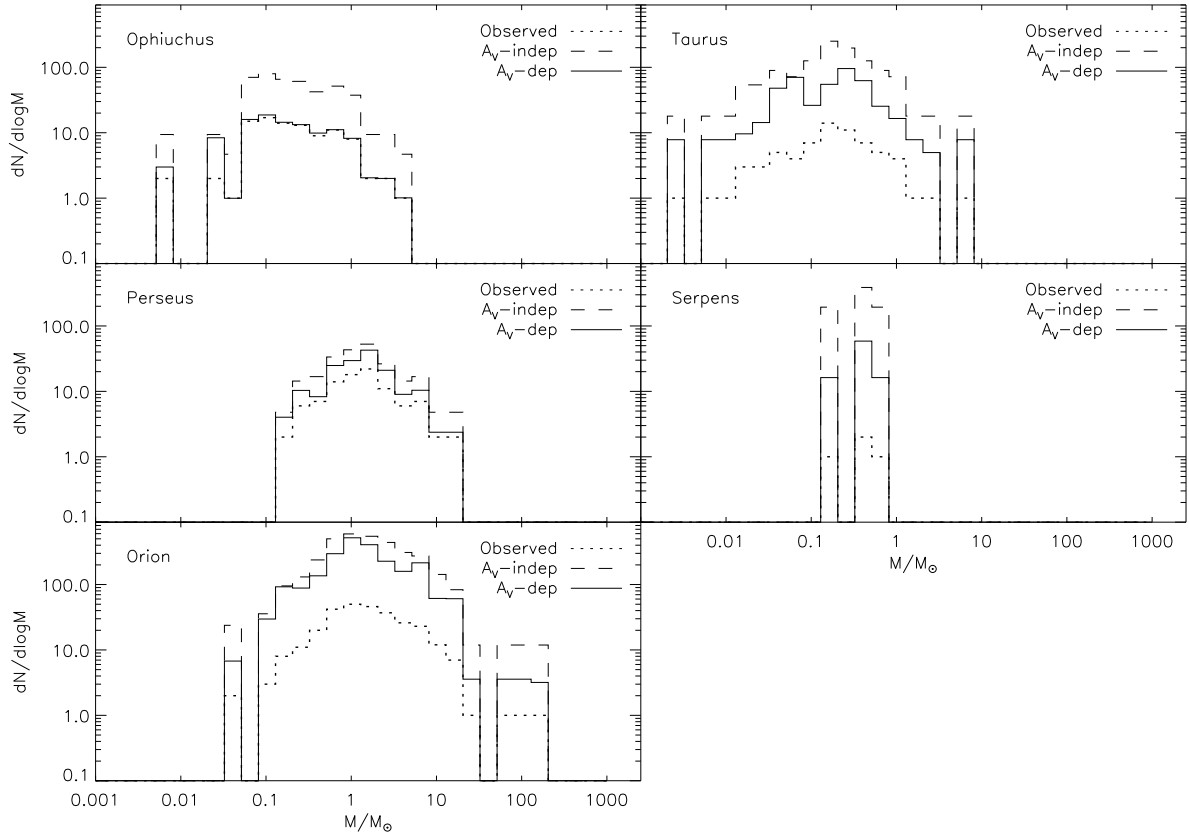}
\caption{Observed starless CMFs are shown as dotted histograms for all five clouds with the predicted $A_V$-dependent starless CMFs as solid histograms and predicted $A_V$-independent starless CMFs as dashed histograms. Extinction ranges for the $A_V$-dependent starless CMFs of all clouds were $\Delta A_V = 4$. For Serpens, we considered the entire extinction map region to produce the extrapolated predicted CMFs (see text). \label{predCMFs+obsCMFs}}
\end{figure}
\clearpage

\begin{figure}[h!]
\includegraphics[scale=0.92]{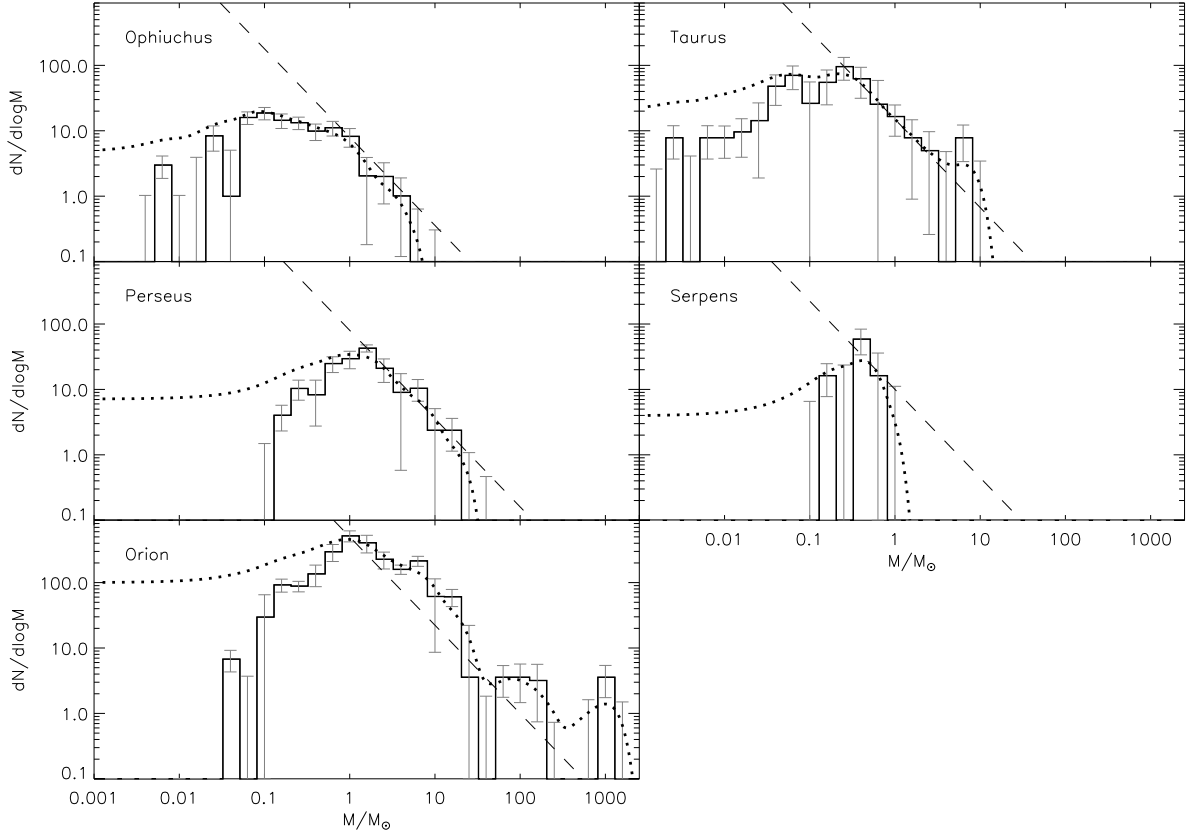}
\caption{Predicted starless CMFs for all five clouds using the $A_V$-dependent extrapolation method. The uncertainties were measured by finding the standard deviation from changes in the temperature. The dashed lines represent a Salpeter slope. Also included is the final distribution obtained from replacing the counts in the mass histograms at each $A_V$ range by a Gaussian (dotted curve). See text for more details. \label{predCMFs+fit+gaus}}
\end{figure}
\clearpage

\begin{figure}[h!]
\includegraphics[scale=0.92]{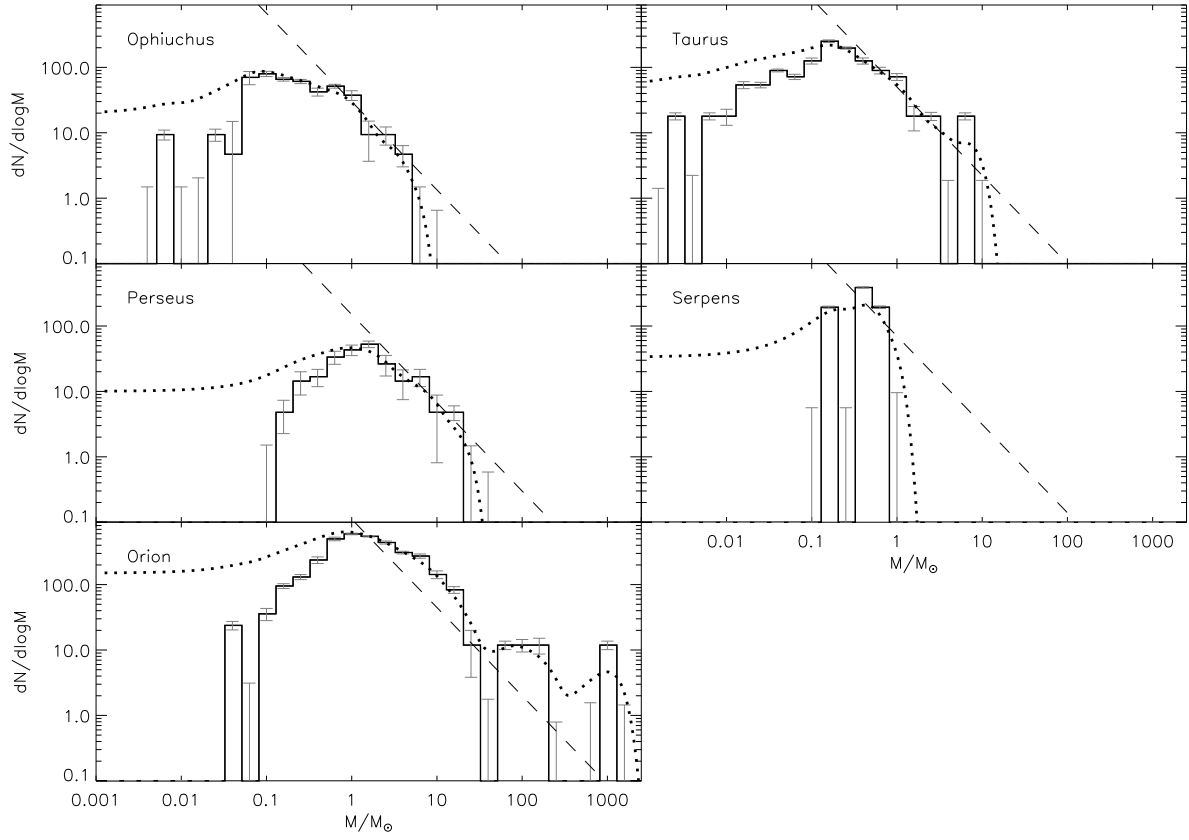}
\caption{Same as Figure \ref{predCMFs+fit+gaus} but for the $A_V$-independent extrapolation. \label{predPixCMFs+fit+gaus}}
\end{figure}
\clearpage


\begin{table}[h]
\caption{Cloud Properties\label{cloudMass}}
\begin{tabular}{lclcc}
\hline
Cloud  & D (pc) & Reference\tablenotemark{a} & M (\Msun) & Area (deg$^2$)\\
\hline\hline
Ophiuchus & 125 & \citealt{Enoch08young} & 1.0 x 10$^4$ & 67\\
Taurus & 140 & \citealt{Goldsmith08}& 3.9 x 10$^4$ & 250\\
Perseus & 250 & \citealt{Enoch08young}& 3.3 x 10$^4$ & 66\\
Serpens\tablenotemark{b} & 260 &  \citealt{Enoch08young}& 1.9 x 10$^4$ & 12.3\\
Orion & 450 & \citealt{PetersonMegeath08}& 2.6 x 10$^5$ & 147\\
\hline
\end{tabular}
\tablenotetext{a}{Reference for our assumed distances.}
\tablenotetext{b}{Since the boundary between the Serpens and Auriga molecular clouds is unclear, we re-measured the mass of Serpens using the cloud boundary from the Herschel Gould Belt survey (see http://starformation-herschel.iap.fr/gouldbelt/gouldbelt\_stage2.pdf), and we obtained a mass of 1.0 x 10$^4$ \Msun\ over 6.3 deg$^2$.}
\end{table}

\begin{table}[h]
\caption{Assumed Properties\label{prop}}
\begin{tabular}{lcl}
\hline
Cloud  & $T_d$ (K) & Reference\tablenotemark{a} \\
\hline\hline
Ophiuchus & 15 & \citealt{Friesen09} \\
Taurus & 13 & \citealt{Andre00} \\
Perseus & 11 & \citealt{Rosolowsky08} \\
Serpens &  17 & \citealt{Schnee05} \\
Orion & 20 & \citealt{Wilson99} \\
\hline
\end{tabular}
\tablenotetext{a}{Reference for our assumed temperatures.}
\end{table}

\begin{table}[h!]
\caption{Area Observed by Each Survey}\label{pixels}
\begin{tabular}{lccc}
\hline
\multirow{2}{*}{Cloud} & SCUBA & Spitzer & 2MASS\tablenotemark{a}\\
& (deg$^2$) & (deg$^2$) & (deg$^2$)\\
\hline\hline
Ophiuchus &  2.4 &  6.6\tablenotemark{b}  & 102\\  
Taurus & 0.94 &  44\tablenotemark{b}  &  327\\
Perseus &  2.7  & 3.9\tablenotemark{b}  &  99.8\\
Serpens & 0.05  & 0.85\tablenotemark{b}  &  12.3\\
Orion &  1.4 &  9.95\tablenotemark{c} & 214\\ 
\hline
\end{tabular}
\tablenotetext{a}{Areas of the entire 2MASS maps. For Ophiuchus and Orion, the 2MASS maps were edited to remove the Scorpius and Monoceros clouds, respectively (see Figure \ref{complete}).}
\tablenotetext{b}{Area with both MIPS and IRAC data.}
\tablenotetext{c}{Area with complete 4-band IRAC coverage. MIPS observations nearly covered the same regions.}
\end{table}

\begin{table}[h!]
\caption{Summary of Cuts to the SCUBA Core Candidate List\label{cuts}}
\begin{tabular}{lccccc}
\hline
Cloud &  Initial\tablenotemark{a} & $A_V > 3$ & $S_{850} > 0$\tablenotemark{b} & Visual & $S_{peak} >$ 0.15 Jy beam\negExp \\
\hline\hline
Ophiuchus & 151 & 150 & 150 & 134 & 124\\
Taurus & 172 & 162 & 147 & 117 & 87 \\
Perseus & 246 & 232 & 231 & 170 & 143 \\
Serpens & 19 & 19 & 19 & 15 & 15 \\
Orion & 448 & 436 & 431 & 392 & 375\\
\hline
\end{tabular}
\tablenotetext{a}{Number of submillimeter core candidates within our 2MASS extinction maps.}
\tablenotetext{b}{All core candidates from the SLC were initially identified using a Clumpfind threshold of 3$\sigma$, where $\sigma$ is the noise level in each map (see \S \ref{ScubaDataSection}). We used the alternative flux, however, which measured the flux for each core again, but with a Clumpfind threshold of 3 x 30 mJy beam\negExp, where 30 mJy beam\negExp\ represents the average noise in all the SLC maps. In low-noise maps, the peak 850 $\mu$m flux can be $<$ 90 mJy beam\negExp, and thus, an alternative flux cannot be obtained.}
\end{table}

\begin{table}[h!]
\caption{Remaining Objects After Each Cut\label{myTechIRCuts}}
\begin{tabular}{ccccc}
\hline
Cloud & CC1-2\tablenotemark{a} & S$_{850} > 90$ & CC3 & CC4a-b\\
\hline\hline
Ophiuchus & 3474 & 146 & 116 & 73\\
Taurus & 91771 & 167 & 160 & 33\\
Perseus & 3429 & 159 & 133 & 99 \\
Serpens & 1443 & 34 & 32 & 25\\
Orion & 18908 & 803 & 768 & 226\\
\hline
\end{tabular}
\tablenotetext{a}{CC1 and CC2 for Ophiuchus, Perseus and Serpens included S/N $\ge 5$ for either 24 or 70 $\mu$m and that neither were bandfilled. Orion and Taurus were not bandfilled, and we used S/N $\ge 5$ for the 24 $\mu$m or for all IRAC bands.}
\end{table}

\begin{table}[h!]
\caption{Comparison of Protostellar and Starless Core Numbers\label{compareProto}}
\begin{tabular}{lccc}
\hline
Cloud & Method & Protostellar & Starless\\
\hline\hline
 \multirow{3}{*}{Ophiuchus} & J\o rgensen & 25 & 99\\
 & Enoch & 33 & 91\\
 & this work & 27 & 97\\
  \hline
 \multirow{3}{*}{Taurus} & J\o rgensen & 24 & 63\\
 & Enoch & $\cdots$ & $\cdots$\\
 & this work & 18 & 69\\
 \hline
\multirow{3}{*}{Perseus} & J\o rgensen & 42 & 101\\
 & Enoch & 49 & 94\\
 & this work & 46 & 97\\
 \hline
 \multirow{3}{*}{Serpens} & J\o rgensen & 8 & 7 \\
 & Enoch & 7 & 8 \\
 & this work & 11 & 4\\
 \hline
  \multirow{3}{*}{Orion} & J\o rgensen & 109 & 266\\
  & Enoch & $\cdots$ & $\cdots$\\
 & this work & 83 & 292\\
 \hline
\end{tabular}
\end{table}

\begin{table}[h!]
\caption{Mean Best Fit Slope\label{bestFitResults}}
\begin{tabular}{lcl}
\hline
Cloud & Slopes\tablenotemark{a} & Mass Ranges\tablenotemark{b} (\Msun)\\
\hline\hline
Ophiuchus & $-1.26 \pm 0.20$ & $0.3 < M < 5$; $0.5 < M < 5$; $0.8 < M < 5$\\
Taurus & $-1.22 \pm 0.06$ & $0.3 < M < 3$; $0.5 < M < 3$\\
Perseus & $-0.95 \pm 0.20$ & $1.0 < M < 20$; $1.2 < M < 20$; $1.6 < M < 20$\\
Orion & $-1.69 \pm 0.72$ & $1.4 < M < 32$; $2.0 < M < 32$; $3.2 < M < 32$\\
Orion (no OMC-1) & $-0.93 \pm 0.18$ & $1.4 < M < 20$; $2.0 < M < 20$; $3.2 < M < 20$\\
\hline
\end{tabular}
\tablenotetext{a}{The slopes quoted for the clouds were calculated from ordinary linear regression with each point weighted by the $\Delta T$ standard deviation method (see text). The slope uncertainties come from the linear fitting algorithm.}
\tablenotetext{b}{Our best fit slopes were calculated using \emph{sixlin} developed by \citet{IsobeSixlin}, and we constrained the fits to have at least 4 points of reference.  Serpens is not included because there were not enough mass bins to calculate a least squares fit.}
\end{table}

\begin{table}[h!]
\caption{Comparison of High-Mass Starless Cores\label{compareHighMass}}
\begin{tabular}{lrcc}
\hline
Cloud & Method & Mass Range & Starless\\
\hline\hline
 \multirow{3}{*}{Ophiuchus} & J\o rgensen & $M > 0.5\ \Msun$ & 33\\
 & Enoch & $M > 0.5\ \Msun$ & 25  \\
 & this work & $M > 0.5\ \Msun$ & 27 \\
  \hline
\multirow{3}{*}{Perseus} & J\o rgensen & $M > 1.2\ \Msun$ & 88\\
 & Enoch & $M > 1.2\ \Msun$ & 80 \\
 & this work & $M > 1.2\ \Msun$ & 85\\
 \hline
\end{tabular}
\end{table}

\begin{table}[h!]
\caption{Predicted and Observed Starless Core Numbers\label{predCoreNum}}
\begin{tabular}{lccc}\\[-4mm]
\hline
Cloud& Observed & $A_V$-dependent & $A_V$-independent\\
\hline\hline
Ophiuchus & 97 & $109 \pm 13$ & $455 \pm 46$\\
Taurus & 69 & $467 \pm 87$ & $1239 \pm 149$\\
Perseus & 97 & $165 \pm 18$ & $232 \pm 24$\\
Serpens & 4 & $91 \pm 50$ & $776 \pm 388$\\
Orion & 292 & $2317 \pm 182$ & $3476 \pm 203$\\
\hdashline
Total & 559 & $3149 \pm 350$ & $6178 \pm 810$\\
\hline
\end{tabular}
\end{table}

\begin{table}[h]
\caption{Examples of Starless Cores\tablenotemark{a}}\label{starlessList}
\begin{tabular}{llccccc}
\hline
\multirow{2}{*}{Cloud} & Core Name & $A_V$\tablenotemark{b} & $S_{peak}$\tablenotemark{c} & $S_{850}$\tablenotemark{d} & M\tablenotemark{e} & $R_{eff}$\tablenotemark{f}\\
                         & (J2000) & (mag) & (Jy beam\negExp) & (Jy) & (\Msun) & (pc)\\
\hline\hline    
\multirow{3}{*}{Ophiuchus} & J162538.4-242238 & 10.52 & 0.15 & 0.03 & 0.0073 & 4.1 x 10$^{-3}$\\
 & J162542.8-241708 & 11.77 & 0.15 & 0.30 & 0.0733 & 1.4 x 10$^{-2}$\\
 & J162546.8-241714 & 12.36 & 0.15 & 1.06 & 0.2589 & 2.5 x 10$^{-2}$\\
 \hline 
\multirow{3}{*}{Taurus} & J040443.3+261859 & 8.16 & 0.40 & 0.60 & 0.2355 & 1.7 x 10$^{-2}$\\
 & J040447.9+261923 & 8.16 & 0.34 & 2.93 & 1.1498 & 3.6 x 10$^{-2}$\\
 & J041828.6+282735 & 16.91 & 0.15 & 0.02 & 0.0079 & 4.0 x 10$^{-3}$\\
 \hline
 \multirow{3}{*}{Perseus} & J032544.0+304026  & 11.63 &  0.34  &  0.25 &  0.4279  & 2.0 x 10$^{-2}$\\
& J032703.2+301513   & 5.10  &  0.23   & 0.83  & 1.4207  &   4.1 x 10$^{-2}$\\
& J032829.4+310956   & 4.72  &  0.21   & 0.96  & 1.6433   &  4.5 x 10$^{-2}$\\
\hline
\multirow{3}{*}{Serpens} & J183001.4+010903 & 14.66 & 0.40 & 0.70 & 0.6035 & 2.8 x 10$^{-2}$\\
& J183002.6+010827 & 12.86 & 0.42 & 1.00 & 0.8621 & 3.5 x 10$^{-2}$\\
& J183005.0+011515 & 14.16 & 0.21 & 0.24 & 0.2069 & 2.2 x 10$^{-2}$\\
\hline
\multirow{3}{*}{Orion} & J051938.8-055144 & 7.55 & 0.15 & 0.15 & 0.3020 & 3.4 x 10$^{-2}$\\
& J051940.4-055150 & 8.22 & 0.17 & 0.11 & 0.2215 & 2.9 x 10$^{-2}$\\ 
& J051948.1-055202 & 9.71 & 1.07 & 3.39 & 6.8254 & 9.5 x 10$^{-2}$\\
\hline
\end{tabular}
\tablenotetext{a}{This table is published in its entirety in the electronic edition. A portion is shown here for guidance regarding its form and content. All cores were classified using our technique.}
\tablenotetext{b}{Visual extinction measured towards the location of peak SCUBA 850 $\mu$m flux.}
\tablenotetext{c}{The peak SCUBA 850 $\mu$m flux (at the position of the source).}
\tablenotetext{d}{Total SCUBA 850 $\mu$m flux within the core, where the core was defined by Clumpfind with a threshold of 90 mJy beam\negExp.}
\tablenotetext{e}{Mass derived from total SCUBA 850 $\mu$m flux using Equation \ref{massEq} with the assumed temperatures listed in Table \ref{prop}.}
\tablenotetext{f}{The effective radius of the core determined by the Clumpfind area, A, where $R_{eff} = \sqrt{A/\pi}$ for a Clumpfind threshold of 90 mJy beam\negExp.}
\end{table}

\begin{table}[h]
\caption{Examples of Protostellar Cores\tablenotemark{a}\label{protoList}}
\begin{tabular}{llccccc}
\hline
\multirow{2}{*}{Cloud} & Core Name & $A_V$\tablenotemark{b} & $S_{peak}$\tablenotemark{c} & $S_{850}$\tablenotemark{d} & M\tablenotemark{e} & $R_{eff}$\tablenotemark{f}\\
                         & (J2000) & (mag) & (Jy beam\negExp) & (Jy) & (\Msun) & (pc)\\
\hline\hline    
\multirow{3}{*}{Ophiuchus} & J162540.5-243020 & 14.10 & 0.23 & 1.72 & 0.4201 & 2.8 x 10$^{-2}$\\
 & J162551.5-243256 & 22.52 & 0.15 & 0.27 & 0.0659 & 1.2 x 10$^{-2}$\\
 & J162610.4-242056 & 29.68 & 0.65 & 2.40 & 0.5861 & 2.7 x 10$^{-2}$\\
 \hline 
\multirow{3}{*}{Taurus} & J041354.4+281129 & 24.83 & 0.19 & 0.36 & 0.1412 & 1.6 x 10$^{-2}$\\
 & J041426.2+280601 & 14.23 & 1.03 & 0.85 & 0.3336 & 1.4 x 10$^{-2}$\\
 & J041828.6+282735 & 15.07 & 0.69 & 1.52 & 0.5965 & 2.4 x 10$^{-3}$\\
 \hline
 \multirow{3}{*}{Perseus} & J032522.2+304514  & 24.21 & 1.41 & 2.13 &  3.6460 &  3.9 x 10$^{-2}$ \\
& J032536.1+304514   & 10.11  &  5.46 &  10.09 & 17.2712 & 5.6 x 10$^{-2}$\\
& J041942.3+271337   & 7.54 & 2.23  &  2.88  & 4.9298   &  3.7 x 10$^{-2}$\\
\hline
\multirow{3}{*}{Serpens} & J182948.2+011639 & 27.01 & 4.16 & 13.21 & 11.3883 & 6.2 x 10$^{-2}$\\
& J182949.8+011515 & 25.54 & 9.87 & 15.17 & 13.0708 & 5.5 x 10$^{-2}$\\
& J182951.4+011633 & 17.68 & 1.76 & 3.91 & 3.3708 & 3.8 x 10$^{-2}$\\
\hline
\multirow{3}{*}{Orion} & J053443.8-054126 & 6.85 & 0.48 & 0.51 & 1.0268 & 4.6 x 10$^{-2}$\\
& J053453.0-054138 & 7.07 & 0.19 & 0.49 & 0.9866 & 5.6 x 10$^{-2}$\\ 
& J053502.3-053756 & 9.31 & 1.01 & 3.67 & 7.3892 & 7.8 x 10$^{-2}$\\
\hline
\end{tabular}
\tablenotetext{a}{This table is published in its entirety in the electronic edition. A portion is shown here for guidance regarding its form and content. All cores were classified using our technique.}
\tablenotetext{b}{Visual extinction measured towards the location of peak SCUBA 850 $\mu$m flux.}
\tablenotetext{c}{The peak SCUBA 850 $\mu$m flux (at the position of the source).}
\tablenotetext{d}{Total SCUBA 850 $\mu$m flux within the core, where the core was defined by Clumpfind with a threshold of 90 mJy beam\negExp.}
\tablenotetext{e}{Mass derived from total SCUBA 850 $\mu$m flux using Equation \ref{massEq} with the assumed temperatures listed in Table \ref{prop}.}
\tablenotetext{f}{The effective radius of the core determined by the Clumpfind area, A, where $R_{eff} = \sqrt{A/\pi}$ for a Clumpfind threshold of 90 mJy beam\negExp.}
\end{table}

\end{document}